**Carbon Nitride: Physical properties and Applications**

Shilpi Kumari[1], Soubhagyam sharma[2], Manisha Kumari[1], Manish Kumar Singh[1], Rakesh K. Prasad[3], Kwang-geol Lee[3], and Dilip K. Singh[1]*

[1]Department of Physics, Birla Institute of Technology Mesra, Ranchi, India-835215.

[2]Department of Physics, Ulsan National Institute of Science and Technology, Korea.

[3]Department of Physics, Hanyang University, Korea.

*Email: dilipsinghnano1@gmail.com

## Abstract

Graphitic carbon nitride (g-$C_3N_4$) has emerged as a versatile, metal-free semiconductor with applications spanning over broad range of domains encompassing energy storage, environmental remediation and sensing. Despite significant progress in recent years, there remains a lack of comprehensive discussion on the g-$C_3N_4$'s evolving role in next-generation technologies and the engineering strategies needed to overcome existing challenges. In this review article, the critical assessment of the physicochemical properties of g-$C_3N_4$ which holds potential to enable its function across diverse applications has been elucidated. Current advances in doping, heterojunction formation and composite engineering that enhances its catalytic and electronic performance has been summarized. The article also presents future research directions to unlock the full potential of g-$C_3N_4$ as a useful material in sustainable and intelligent systems.

## Introduction

Ever since the discovery of graphene in 2004 **[1]**, there has been a huge increase in the research and development of carbon based lower-dimension materials **[2-5].** This is due to the abundant and cheap availability of carbon as an element. Another reason is the organic and environment friendly nature of carbon-based nano materials, which can be used as resource for the production of energy in a non-toxic manner **[6-7]**. Carbon nitrides have been pretty promising carbon-based material complimenting harmful carbon-based compounds in material applications. Among all of the carbon nitride allotropes, g-$C_3N_4$ is regarded as the most thermally and chemically stable compound under ambient conditions **[8].** Melon, which is one of the parent polymers which decompose to form g-$C_3N_4$, is considered to be one of the oldest polymers with its synthesis and nomenclature dating around 1834 **[9].** The compound could not be studied further up until recently due to its high chemical stability as well as insolubility in acidic, basic and organic solutions.

g-$C_3N_4$ is a polymeric, highly defective and nitrogen rich metal free material. It's carbon and nitrogen composition along with its organic chemistry makes possible to modify its reactivity without much change in its chemical composition **[10]**. Due to its configuration, a number of organic, inorganic compounds or, metals can be combined with g-$C_3N_4$ to form composites having desirably tuned properties **[11-13]**. g-$C_3N_4$ is a medium band gap, indirect semiconductor having a band gap of 2.7 eV which along with other properties such as good visible light absorption and high photoluminescence makes it suitable for optoelectronics **[14]**, electrochemistry **[15]** as well as sensing applications **[16-17].** The fluorescence properties of g-$C_3N_4$ have been utilized for bio-imaging purposes in order to identify internal damages/ diseases inside the body. **[18]**. Recently, various g-$C_3N_4$ composites have shown remarkable results in energy storage applications such as Li-ion storage **[19]** and supercapacitors **[20]**.

The cost effectiveness, facile synthesis, robust nature plus the chemical and electronic configurations of g-$C_3N_4$ has steered a wider pathway for research in organic semiconductor electronics. This review article presents an account of basics of synthesis of g-$C_3N_4$ in bulk and nano-dimensions in addition to the structure, properties and its major applications.

**Synthesis of Bulk g-$C_3N_4$**

The popular methods used for the bulk synthesis of graphitic $C_3N_4$ are namely solvothermal method, solid state reaction or thermal nitration depending upon the desired crystallinity or C: N ratio. The easiest and most common method is the direct heating of reagents in air or in the presence of inert atmosphere. These reagents are nitrogen rich compounds having C-N bonded molecules as their core configuration. These structures exist in the form of triazine ($C_3H_3N_3$) isomers, mostly- 1, 3, 5- triazine or as heptazine/ tri-s-triazine (having three triazine rings fused together) as shown in **Fig. 1.** However, the compounds formed by tri-s-triazine based reagents have been found to be 30 $kJmol^{-1}$ more stable. **[21]**. Previously, the common reagents used for the bulk synthesis were cyanamide and/or dicyandiamide which went through polycondensation to form various intermediates depending on the change in temperature.

Deamination occurs at temperatures above its melting point to give melamine-based compounds up to 350 °C. Melamine further rearranges to form tri-s-triazine based compounds at 390 °C. These polymers condense to form $C_3N_4$ around 520 °C **[8].** Since cyanamide is costly as well as highly explosive, we simply use melamine or urea as precursors. On further heating, Melamine sublimates and condensates in the temperature range of 280-400 °C into melem and melon as shown in **Fig. 2**. At temperatures above 650 °C the g-$C_3N_4$ starts becoming unstable **[23].** We can also synthesize bulk g-$C_3N_4$ using urea as a precursor. *Liu et al.* thermally treated urea in a closed crucible under ambient pressure and in the presence of air. Urea when heated at around 550 °C - 660 °C gives yellowish residue which is g-$C_3N_4$ **[24].** The closed environment prevents sublimation of the precursors. Also, the temperature range of 550 °C to 600 °C yields g-$C_3N_4$ having different performances depending upon the temperature range and the time period for which it is kept at a particular temperature. *Papailias et al.* studied the differences of processing temperature on the structure and photocatalytic properties of g-$C_3N_4$. They observed that the final products were different in color and the final product yield decreased with the increase in temperature as well **[25].**

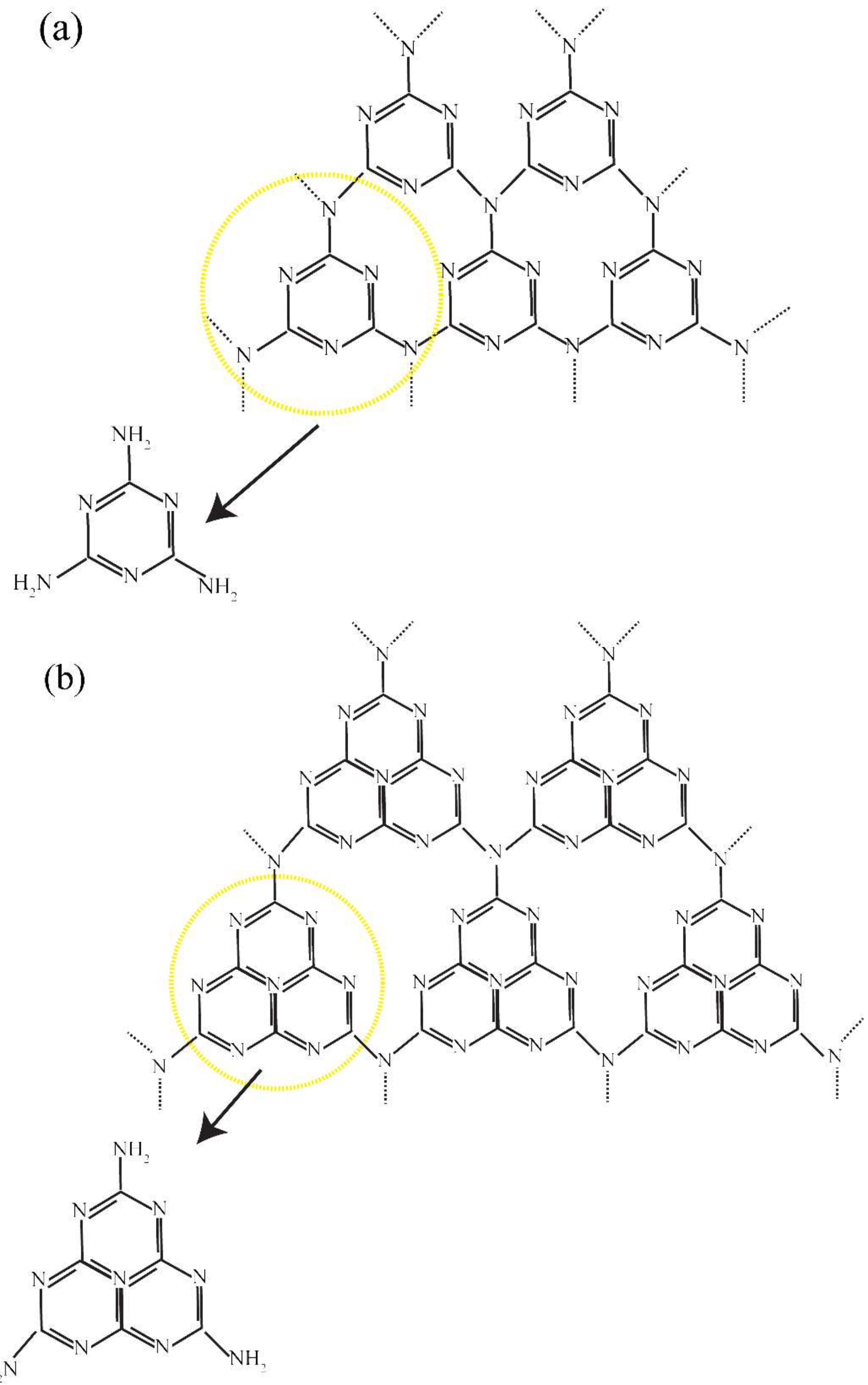

**Figure 1** (a) Triazine structure and (b) tri-s-triazine (heptazine) structure of g-$C_3N_4$ **[22].**

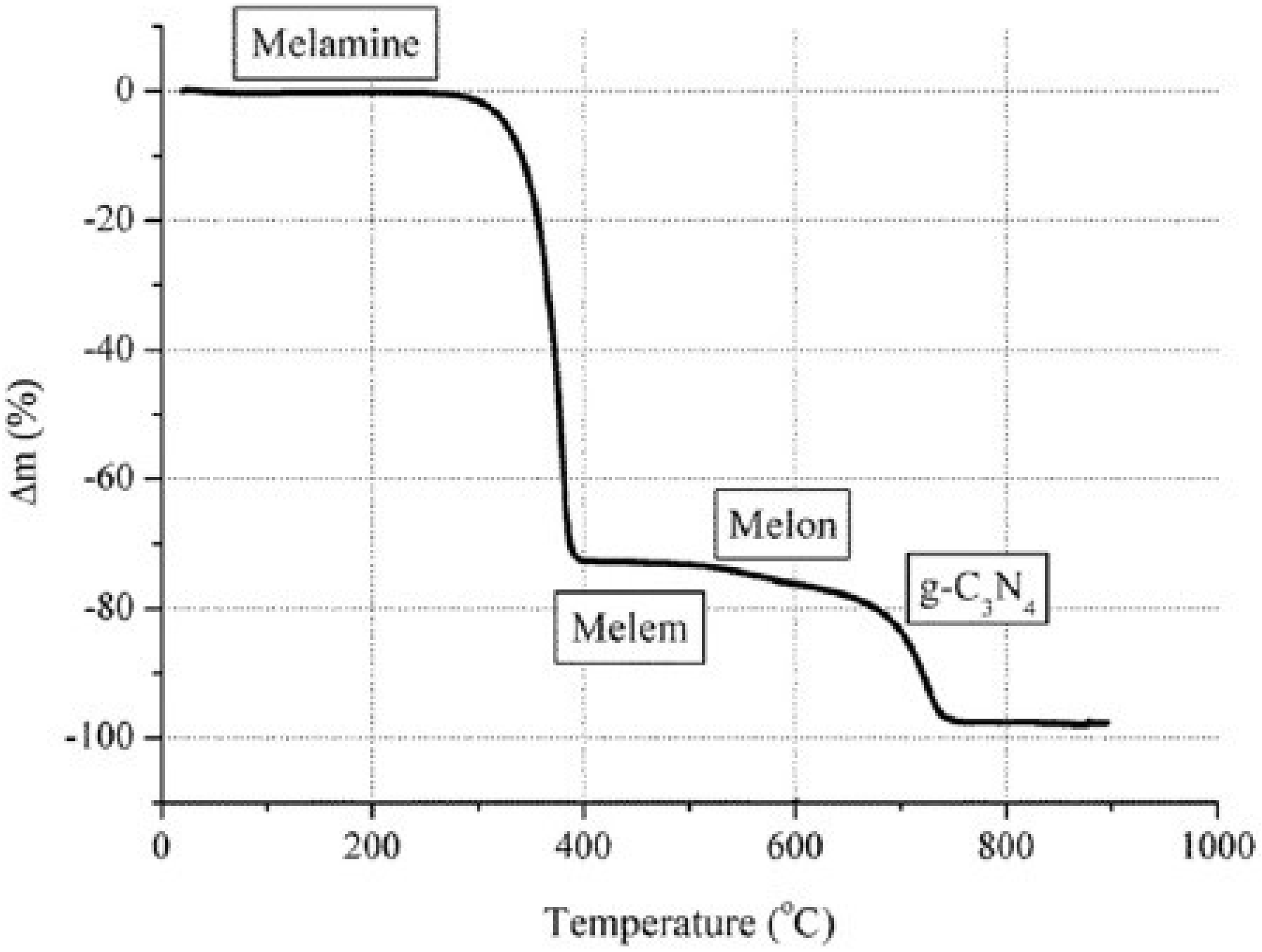

Fig. 2 Thermogravimetric curve of annealing of melamine leading to formation of g-$C_3N_4$ at temperature of ~700 °C. **[23].**

**Nano/Micro Dimensional Synthesis**

Due to the extraordinary performance of g-$C_3N_4$ in electronic as well as electrochemical fields, the need to synthesize it easily in nano or micro dimensions in order to make hetero-structures has increased. The growth of thin films via Physical or Chemical vapor deposition has also been an issue. *Tanaka et al.* used ion beam assisted deposition using g-$C_3N_4$ pellets for thin film growth on Si substrate but the films were of C-$N_x$ and had presence of hydrogen as well **[26].** In the thin film deposition techniques, the quantization of nitrogen has not been successfully done yet. Popular methods to reduce the dimensions of bulk g-$C_3N_4$ are chemical exfoliation and thermal oxidation. *Feng et al.* used Hummers method to form g-$C_3N_4$ nanorods from bulk g-$C_3N_4$ **[27].** Pawar et al. synthesized g-$C_3N_4$ microrods using different concentrated acids to etch and exfoliate the bulk g- $C_3N_4$ simultaneously. They found out that the rods formed using $H_2SO_4$ were the most uniform **[28].** *Tahir et al.* annealed melamine with appropriate amount of ethanol and $HNO_3$ in a CVD to obtain nanofibers having diameter of 100 nm and 20 μm in length **[29].** *Wang et al.* synthesized nanotube type g-$C_3N_4$ by directly heating packed melamine **[30].** The chemical exfoliation of bulk g-$C_3N_4$ is most widely used to obtain nanosheets as shown in **Fig. 3.** *Zhang et al.* calculated and compared the surface energies of bulk g-$C_3N_4$ and $H_2O$ to find that g-$C_3N_4$ nanosheets can be exfoliated using water. The thickness of the nanosheets were found to be ~2.5 nm **[31-32].** *Xu et al.* in the same year, extracted a single atomic layer nanosheet with thickness ~ 0.4 nm using chemical exfoliation **[33].** *Niu et al.* synthesized g-$C_3N_4$ nanosheets by thermal 'oxidation' etching in the presence of air and observed that the thickness of g-$C_3N_4$ decreased with increased etching time **[34].** Generally the etching is done with $H_2SO_4$, ethanol, $HNO_3$ or acetone sometimes assisted with water through ultrasonication **[35]**. However, *Fang et al.* obtained the nanosheets having large area aspect ratio i.e. lateral size more than 15 μm using Anhydrous Ethylenediamine as assistant reagent and optimizing the etching treatment time **[36].** The Schematic of synthesis of bulk g-$C_3N_4$ from standard pyrolysis of Urea is shown in **Fig. 4.** *Li et al.* prepared carbon rich g-$C_3N_4$ nanosheets by the self-modification of polymeric melon units through thermally treating bulk g-$C_3N_4$ in air and in $N_2$ atmosphere successively **[37].** *Sun et al.* used commercially available carbonated beverages to synthesize mesoporous g-$C_3N_4$ nanosheets **[38].** Though it has been tougher to obtain 2D thin films of g-$C_3N_4$ without altering the composition of Nitrogen, there have been some successful productions of thin films. *Mao et al.* extracted g-$C_3N_4$ nanosheets on FTO glass to study their photoelectrochemical properties **[39].** *Jia et al.* deposited g-$C_3N_4$ on both sides of a glass substrate simply by using melamine and urea as precursor and heating them at 550 °C in nitrogen rich atmosphere. The crucible was covered completely with aluminum foil. They also found out that the thickness of the film deposited could be changed by changing the amount of precursors used **[40]**. Even though we have come a long way in the synthesis and optimization of nano/micro scale graphitic-$C_3N_4$, the perfect method for large scale production without major changes in the C-N configuration has to be achieved yet.

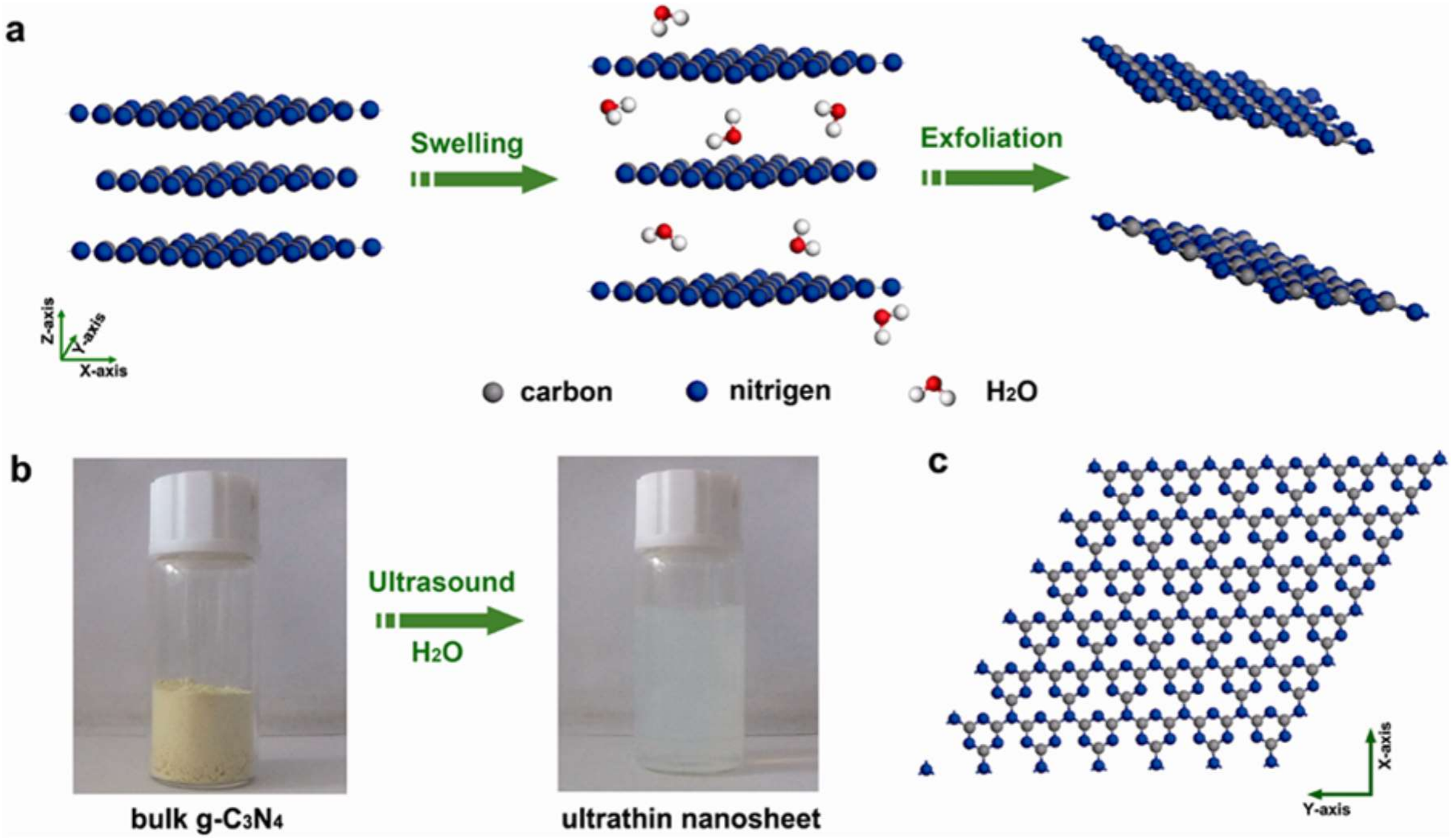

**Fig. 3** Schematic of the synthesis procedure of g-$C_3N_4$. (a) liquid-exfoliation process from bulk g-$C_3N_4$ to ultrathin nanosheets. (b) Photograph of bulk-$C_3N_4$ and suspension of ultrathin g-$C_3N_4$ nanosheets. (c) Schematic of the perfect crystal structure of the g-$C_3N_4$ projected along the z-axis **[31].**

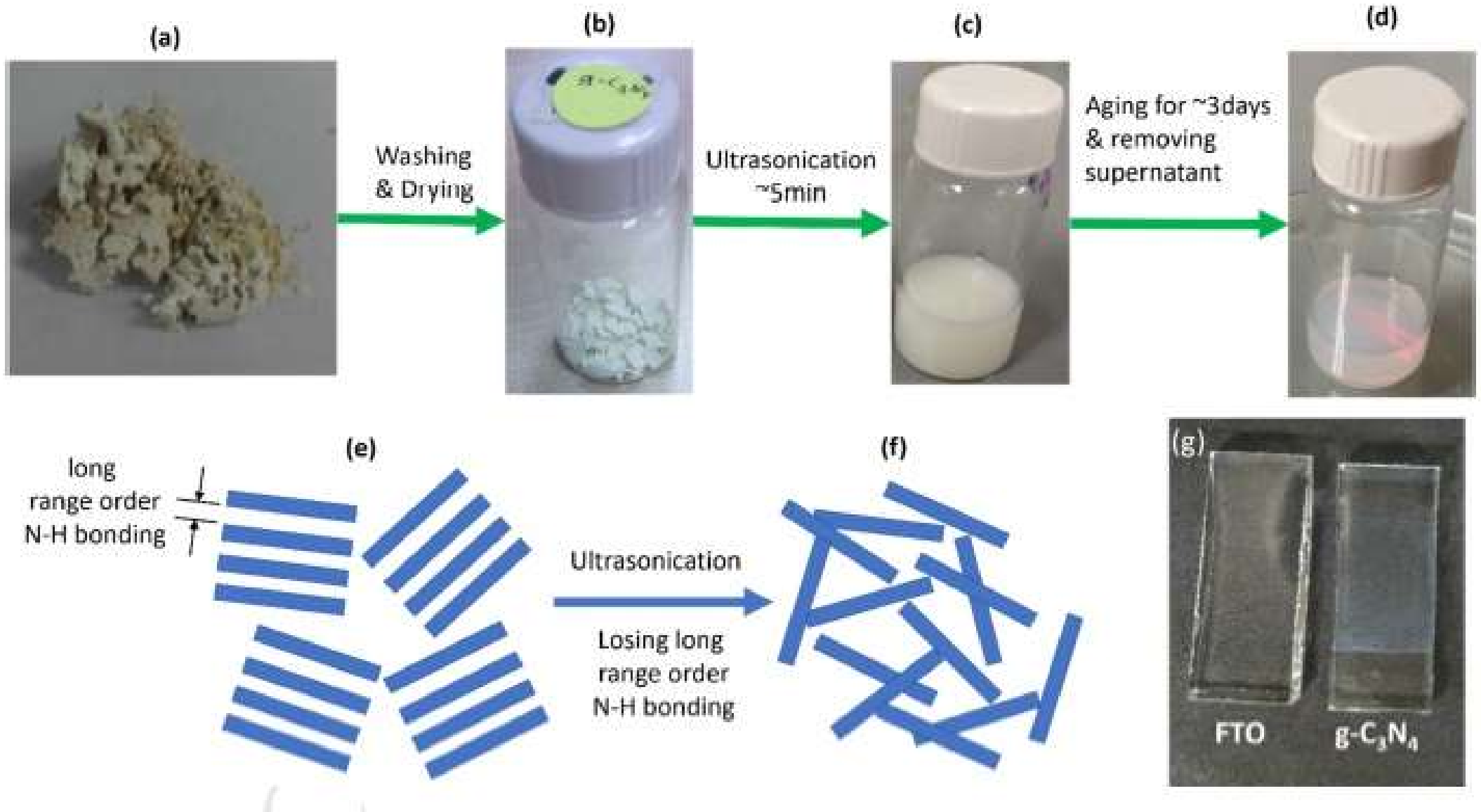

**Figure: 4** Schematic of synthesis of bulk g-$C_3N_4$ (a) standard pyrolysis of Urea (b) Washing / drying and (c) further exfoliation and dispersion in methanol for ~5 min (d) further aging for ~3 days and removing of top supernatant dispersion of highly stable g-$C_3N_4$ nanosheets. (e) Illustration of bulk g-$C_3N_4$, (f) undergoing ultrasonication and resulting in exfoliated g-$C_3N_4$ nanosheets and (g) actual g-$C_3N_4$ thin film together with fluorine doped tin oxide (FTO) glass substrate. **[41]**

## 2. Properties

Graphitic carbon nitride is a polymeric compound with different C-N bonds. In some cases, presence of hydrogen is also observed which indicates the presence of amines due to incomplete decomposition of precursors. It is stable hydrothermally i.e., it is insoluble in water, toluene, ether, THF etc. which makes it a promising candidate for catalysis. Several researchers have characterized graphitic carbon nitride in its bulk as well as exfoliated form to study its optical, electronic as well as electrochemical properties. Besides SEM, TEM and AFM used for the morphological characterization, the characterization techniques to quantify the properties include: thermal gravimetric analysis (TGA) for thermal stability, FTIR spectroscopy, X-ray diffractometry (XRD), X-ray photoelectron spectroscopy (XPS), Raman Spectroscopy as well as Diffuse reflectance spectroscopy (DRS) and Photoluminescence (PL) for the emission and absorption spectra. Here, first the bulk properties of g-$C_3N_4$ would be discussed and then these properties will be compared with that of the exfoliated or synthesized g-$C_3N_4$ in smaller dimensions.

**Structural properties**

The crystallographic properties of bulk as well as nano dimensional g-$C_3N_4$ are mainly studied via their TEM (Transmission electron microscopy), HRTEM (High resolution, AFM, and SEM. Fig. 5 is the HRTEM micrograph of the material. The image indicates a close to hexagonal arrangement of the scaffolding tri-s-triazine structure with apparent interplanar distances between 0.58 and 0.65 nm **[8]**. A perfect crystalline material consisting of tri-s-triazine units would reveal a sixfold or threefold symmetry with distances of d = 0.72 nm. The material is rich in defects, i.e. it grows in the form of polymer material and not as a single crystal. Measurements of the inter-layer distance from the HRTEM micrograph reveals a spacing of d = 0.327 nm, which is in good agreement with the XRD data **[8].**

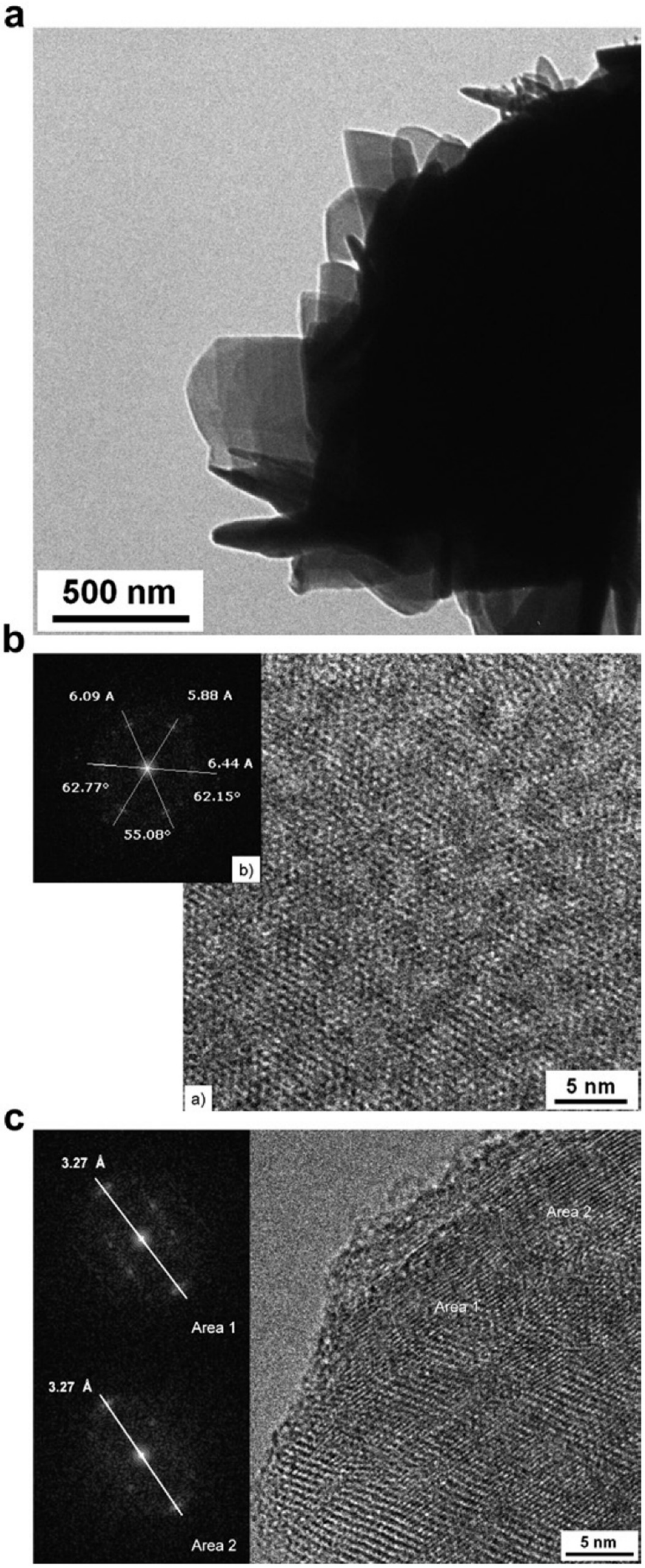

Fig.5 (a) TEM micrograph of the crystallized g-$C_3N_4$. It shows layered platelet like morphology. (b) Shows in-plane organization of the g-$C_3N_4$ formed by tri-s-triazine units. (b) Shows that the hexagonal symmetry is broken, indicating either connectivity defects or angularity of view. The SAED pattern is shown in inset of figure. (c) HRTEM showing Graphite-like stacking of g-$C_3N_4$. Areas 1 and 2 reveal an inter-planar distance of 0.327 nm and a periodicity within the layers. Due to stacking faults streaks are visible **[8].**

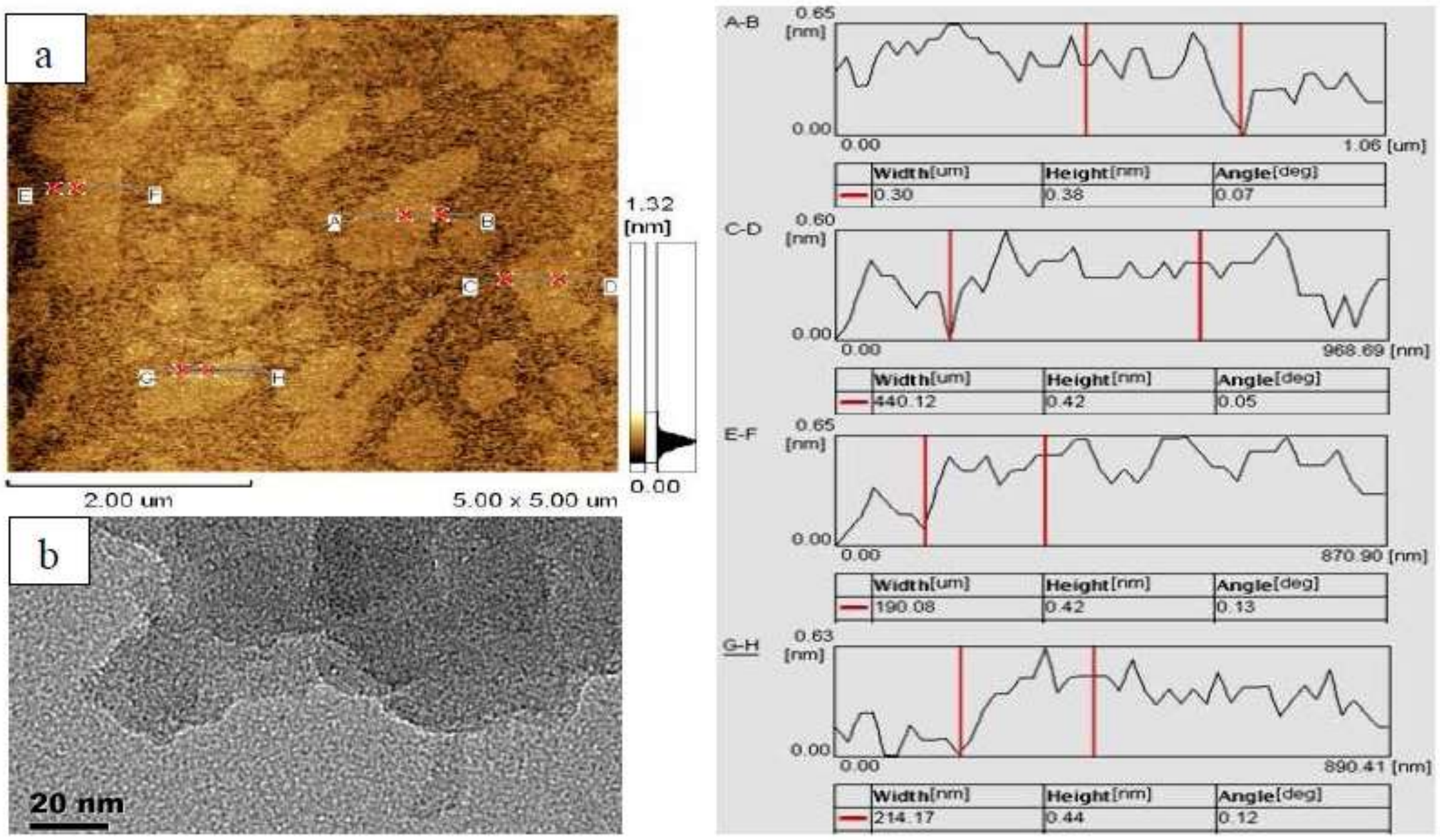

**Fig.6** Typical (a) AFM images and (b) TEM images of the as-prepared Monolayer-$C_3N_4$ nanosheets. The tables and curves in the right panel shows the height and width information of different nanosheets from the AFM images **[33].**

*Xu et al.* exfoliated a single layer of g-$C_3N_4$ and did the morphological analysis via TEM and AFM. The thickness analysis of the nanosheets by AFM revealed an average thickness of about 0.4 nm, which is very close to the theoretical thickness of monolayer g-$C_3N_4$ (about 0.325 nm) as shown in Fig. 6. The very transparent feature of the nanosheets in the TEM analysis indicates the ultrathin thickness of the nanosheets. The darker part in the TEM image can be attributed to the overlap of several g-$C_3N_4$ nanosheets or a multilayer nanosheet **[33].**

The X-Ray Diffraction of bulk g-$C_3N_4$ shows two distinct diffraction peaks, one at ~27.40° indexed as (002) with interlayer spacing d = 0.325 nm of the aromatic units and the other one at ~13.04° indexed as (100) with d = 0.676 nm attributed to interlayer stacking **[42-43]**. The peak at 13.04° vanishes or becomes less pronounced when we characterize nanosheets or nanofibres due to decreased planar size **[44].** When the samples were prepared by direct polymerization, Hong et al. noticed that the (002) peak intensity of as-prepared samples related to the interlayer stacking became weaker and broader with increasing the polymerization temperature, revealing the reduced layer thickness **[45].** In case of nanofibres, Tahir et al. observed only one peak at 27.3° indicating long range inter-planar stacking **[30].** *Niu et al.* also detected that the peak at 27.34° from periodic stacking shifted to 27.67° indicating a decreased gallery distance between the basic sheets in a nanosheet **[35].** It has been observed that due to the formation of crumples and agglomerates, the XRD pattern of exfoliated nanosheets might be similar to that of bulk g-$C_3N_4$ **[46].** *Xu et al.* found that for monolayer g-$C_3N_4$, the peak at 27.04° weakened significantly as visible in Fig. 7 **[33].**

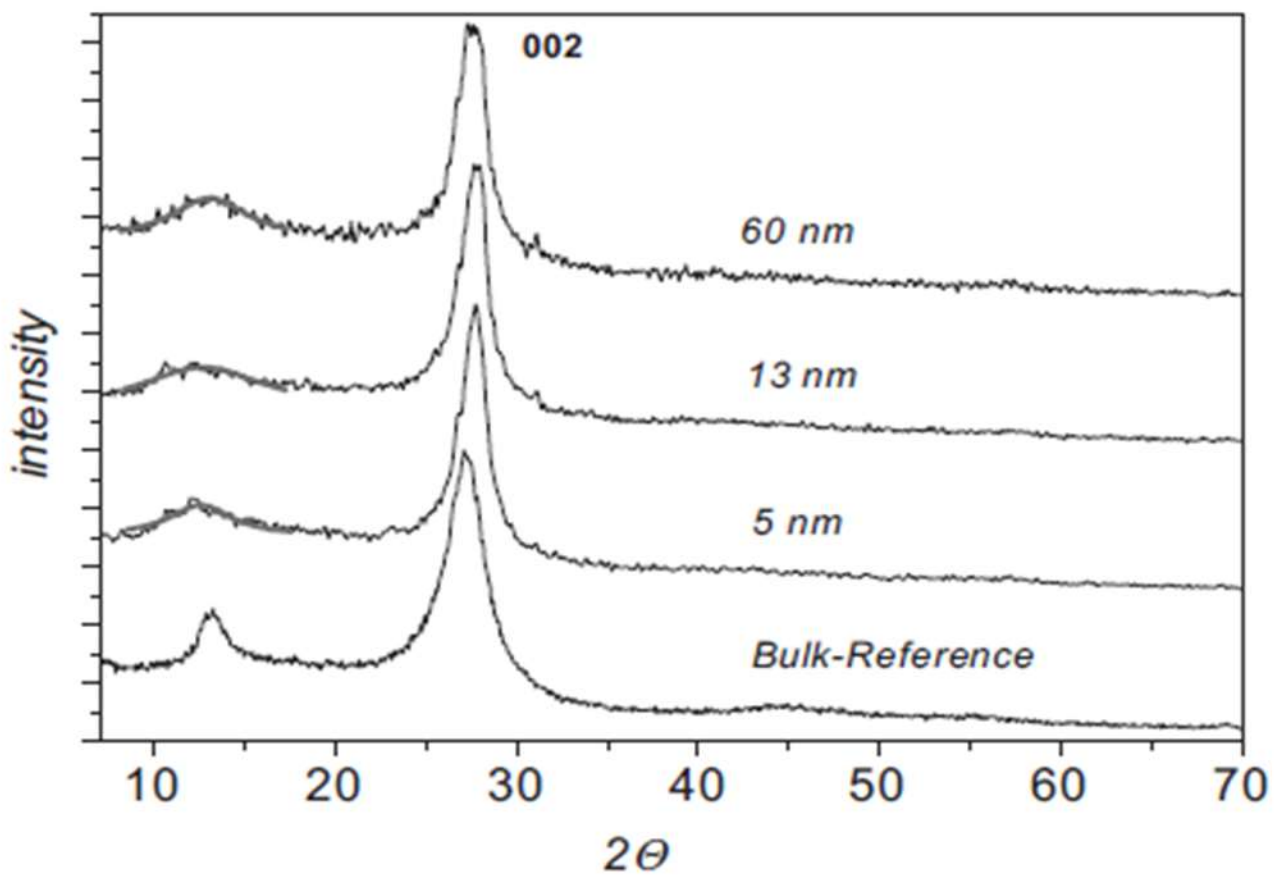

**Fig. 7** X-ray diffractogram of nanoscale $C_3N_4$ with different diameters. The interlayer periodicity (reflex at 2θ ~ 13°, d = 0.68 nm) is broadened due to geometric confinement while, the (002) interlayer spacing is weakly affected **[47].**

Raman Spectroscopy is used to investigate the crystalline quality, symmetry, defects, strain, number of layers and electronics properties of materials **[48]**. The most intense feature is observed in the Raman shift range of 900 to 1800 $cm^{-1}$ with two prominent modes namely D and G-bands. The lower energy one is called the D band and the higher energy one is called the G band. The G mode is related to the main mode of crystalline graphite, the former related to the finite size of the graphitic domains **[49].** The characteristic peaks for bulk g-$C_3N_4$ are observed at 1616, 1555, 1481, 1234, 751, 705, 543, and 479 $cm^{-1}$ **[50].** However, the signature peaks of g-$C_3N_4$ are 711 $cm^{-1}$ (heptazine ring breathing mode) and 1233 $cm^{-1}$ (stretching vibration modes of C=N and C–N heterocycles) as shown in **Fig. 8 [51].** The Raman peaks observed at approximately 707 and 986 $cm^{-1}$ for all the samples and are attributed to different types of ring breathing modes of s-triazine which is present in the g-$C_3N_4$. Generally, the nanosheets show peaks similar to that of the bulk g-$C_3N_4$ i.e. there are two characteristic bands related to the disordered D (around 1300 $cm^{-1}$) and graphitic G (around 1500 $cm^{-1}$) bands, indicating the formation of graphitic structure. The Raman bands at 600 – 800 $cm^{-1}$ can be attributed to the in-plane rotation of sixfold rings in a graphitic carbon nitride layer **[33].** However, the g-$C_3N_4$ nanosheets show a blue shift of about 5 $cm^{-1}$ compared with bulk g-$C_3N_4$, which is due to the phonon confinement effect of exfoliated nanosheets implying their thickness **[52].** *Jiang et al.* observed that the ratios of $I_{726}/I_{685}$ and $I_{616}/I_{555}$ were increased from 0.25 and 4.4 for bulk g-h-heptazine-$C_3N_4$ to 1.0 and 8.1 for 1-layer g-h-heptazine-$C_3N_4$. They also observed that the peaks at 1224 $cm^{-1}$ might be ascribed to the C= C ($sp^2$) bending vibration. It was shifted by 8 $cm^{-1}$ towards high frequency after the bulk sample was exfoliated into the few-layer samples, probably due to electrophilic effect and the quantum effect of the ultrathin (nano-sized) layer coplanar g-h-heptazine-$C_3N_4$, leading to the enhanced strength of C–N covalent bonds **[51].**

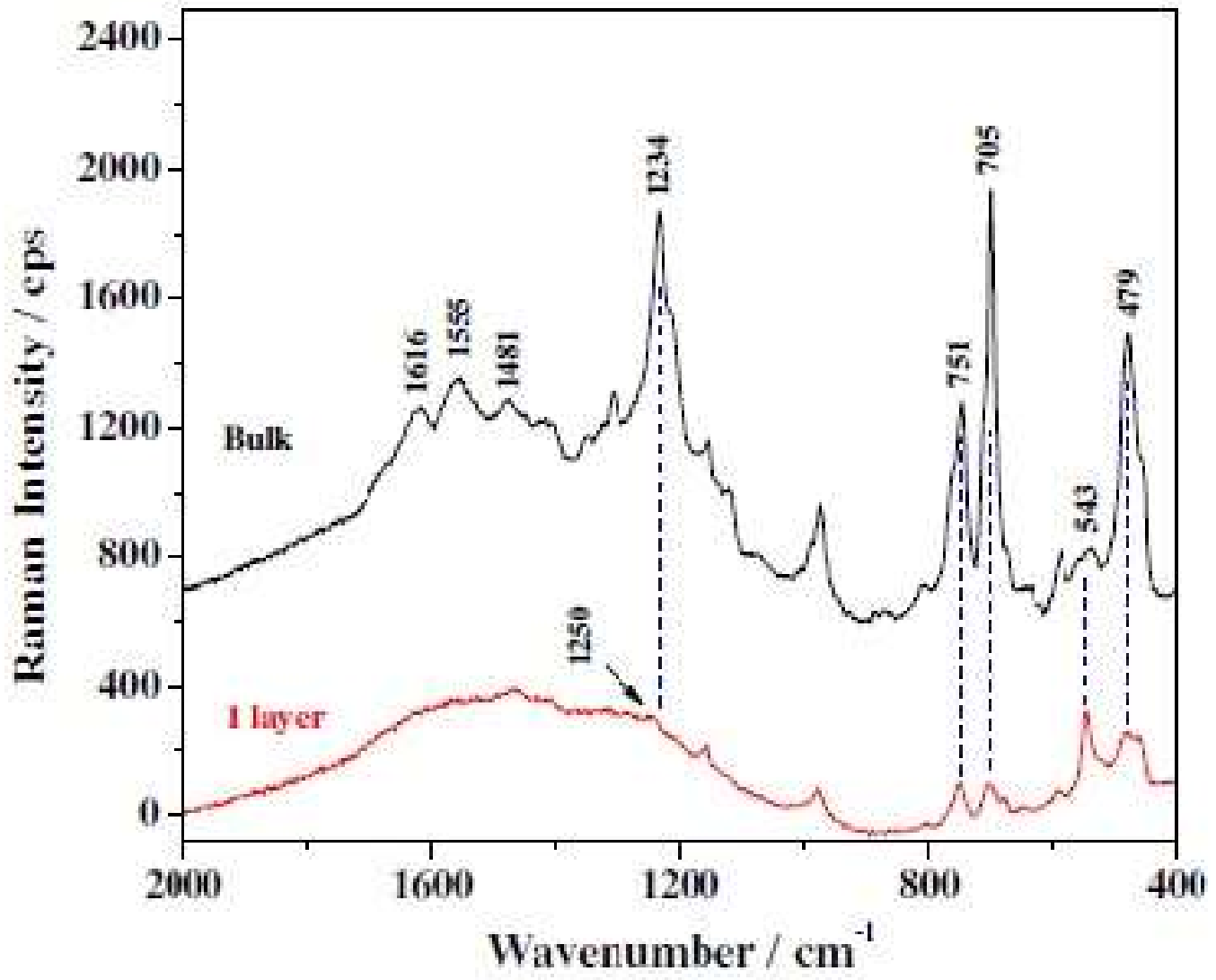

**Fig. 8** Comparison between the Raman spectra of coplanar bulk and 1-layer g-h-heptazine-$C_3N_4$, samples (excitation wavelength 780 nm) **[51].**

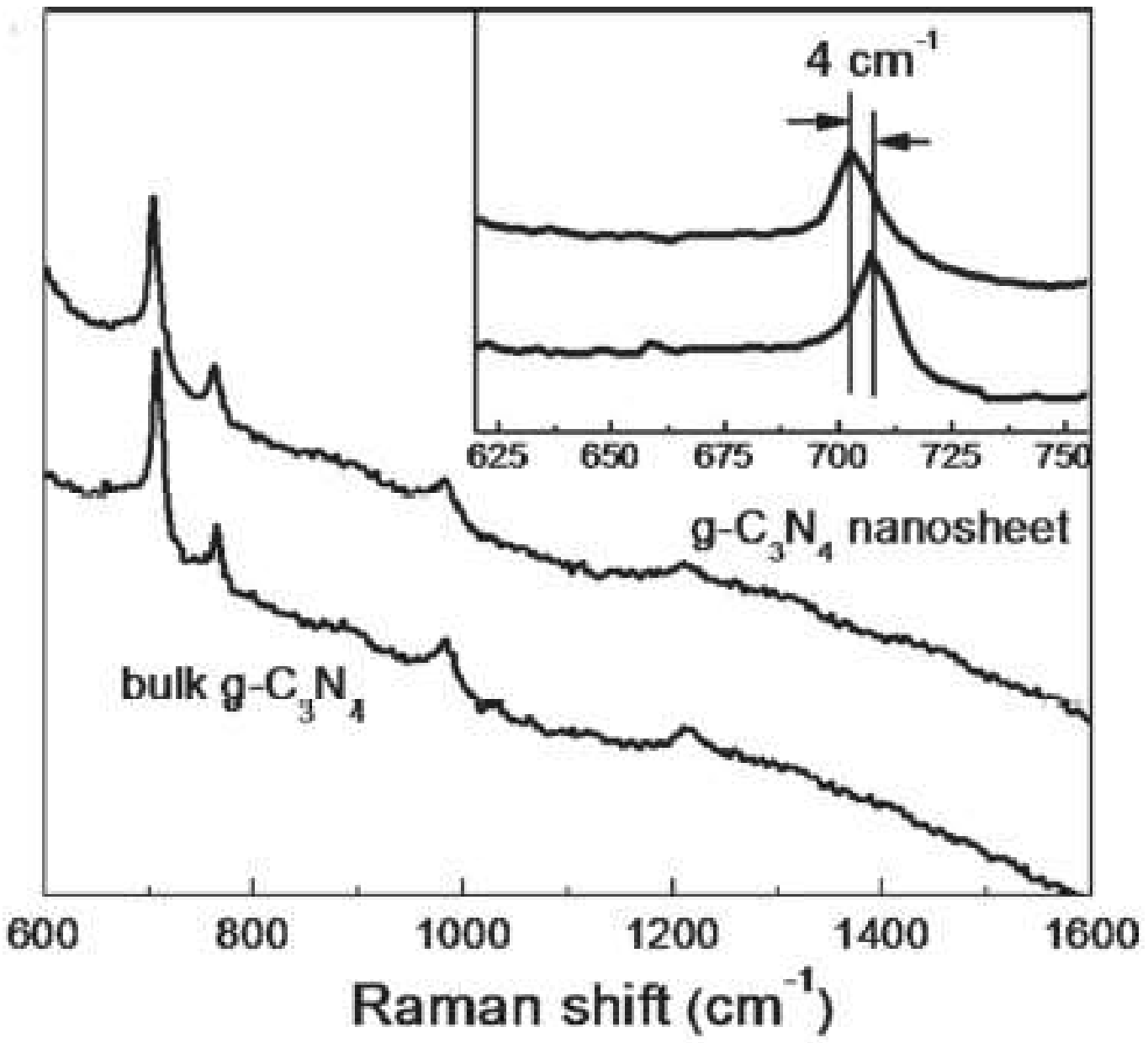

**Fig. 9** Raman spectra of bulk g-$C_3N_4$, and g-$C_3N_4$ nanosheets (upon 785 nm excitation). **[53]**.

The finger prints of various functional groups present on the surface is mapped using Fourier Transform Infrared Spectroscopy (FTIR). In FTIR spectrum, bulk g-$C_3N_4$ shows a sharp peak at 810 $cm^{-1}$, characteristic absorption peak of triazine unit in g-$C_3N_4$ **[40]**. In addition to it, bands are observed in the region of 1570–1634 $cm^{-1}$ and 1258–1480 $cm^{-1}$ typical of vibrations of C-N bonds, which belong to the skeletal stretching modes of aromatic rings **[54-55].** The peak at 1643 $cm^{-1}$ is attributed to C-N stretching vibration modes, while the 1242 $cm^{-1}$, 1322 $cm^{-1}$, 1405 $cm^{-1}$ and 1563 $cm^{-1}$ are associated with aromatic C-N stretching. The prominent absorption bands at 1206 $cm^{-1}$ and 1235 $cm^{-1}$ and possibly 1316 $cm^{-1}$ have been shown to be characteristic of the C-NH-C unit in melam **[56]**. Therefore, a similar structural motif, corresponding to either trigonal CN(-C)C (full condensation) or bridging C-NH-C units (partial condensation) can be inferred for the polymer **[34, 57]**. Broad band between 3100 $cm^{-1}$ to 3300 $cm^{-1}$ shows the uncondensed terminal amino group (-$NH_2$ OR =NH group) which becomes stronger at a higher temperature, i.e. more amino groups are generated **[57].** *Tong et al.* observed that in case of nanosheets, the intensity of peaks 900-1800 $cm^{-1}$ changed with respect to that of the bulk material, resulting from the protonation and disintegration of g-$C_3N_4$ **[44].** *Bai et al.* observed that the peaks were stronger at 1241 $cm^{-1}$ (C-N) and 1631 $cm^{-1}$ (C=N) along with a new peak at 791 $cm^{-1}$ attributed to more heptazine rings and less surface defects for synthesized nanorods **[43]**. *Pawar et al.* observed that the band at 3500 $cm^{-1}$ was strongest for the nanorod sample chemically exfoliated using $H_2SO_4$, which may have resulted from slight oxidation of the g-$C_3N_4$ by the strongly oxidizing sulfuric acid **[28].** *Fang et al.* observed that some peaks of 2D nanosheets were sharper than that of bulk g-$C_3N_4$, which is due to the more ordered packaging of tri-s-triazine motifs in the nanosheets. At 2180 $cm^{-1}$, the vibration peak becomes more distinct which might descend from more $H_2O$ molecules or amino groups bonding on the surface of g-$C_3N_4$, implying that the nanosheets have enlarged open-up surface **[58-59].**

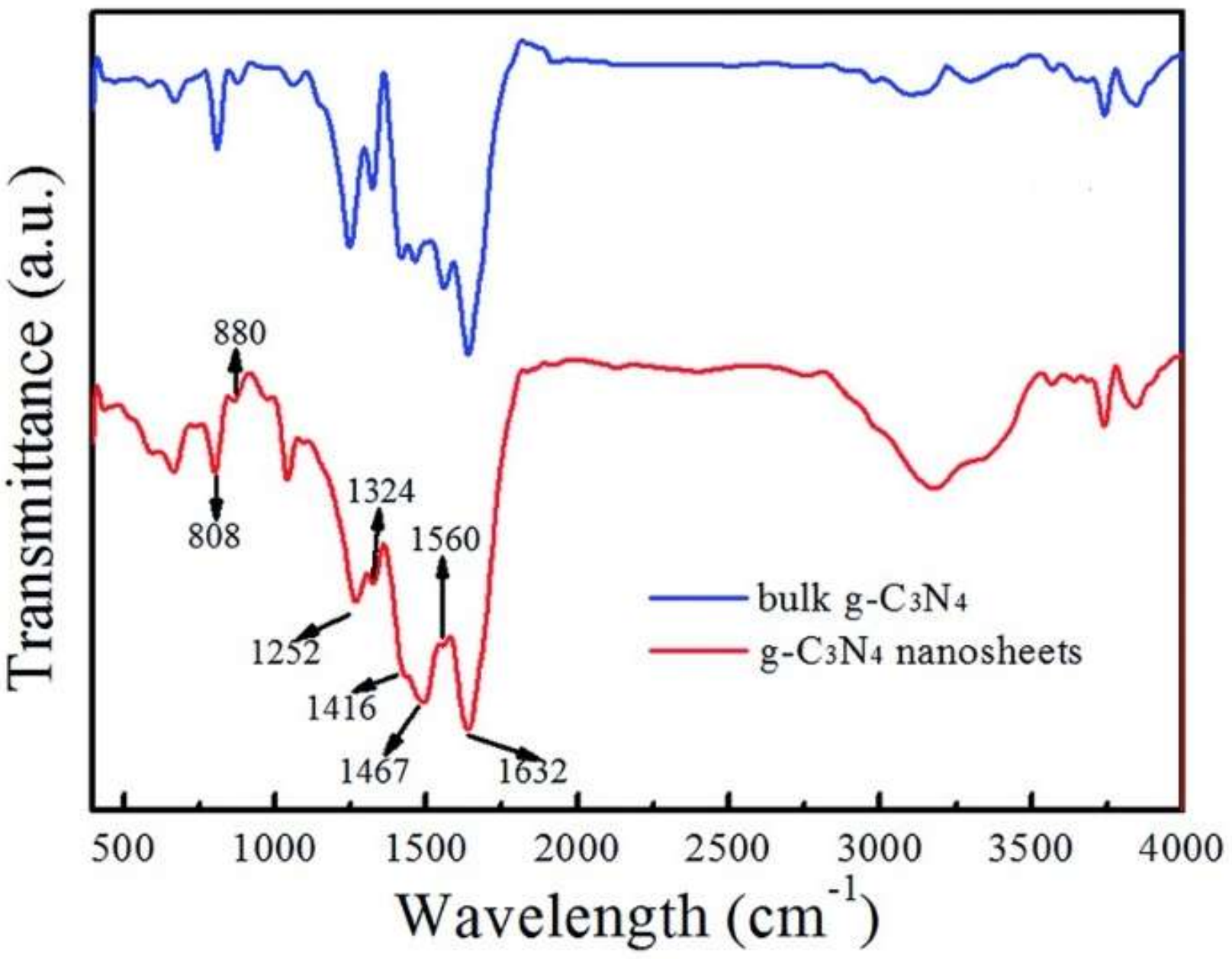

**Fig. 12** FTIR spectra of bulk g-$C_3N_4$ and g-$C_3N_4$ nanosheets **[60].**

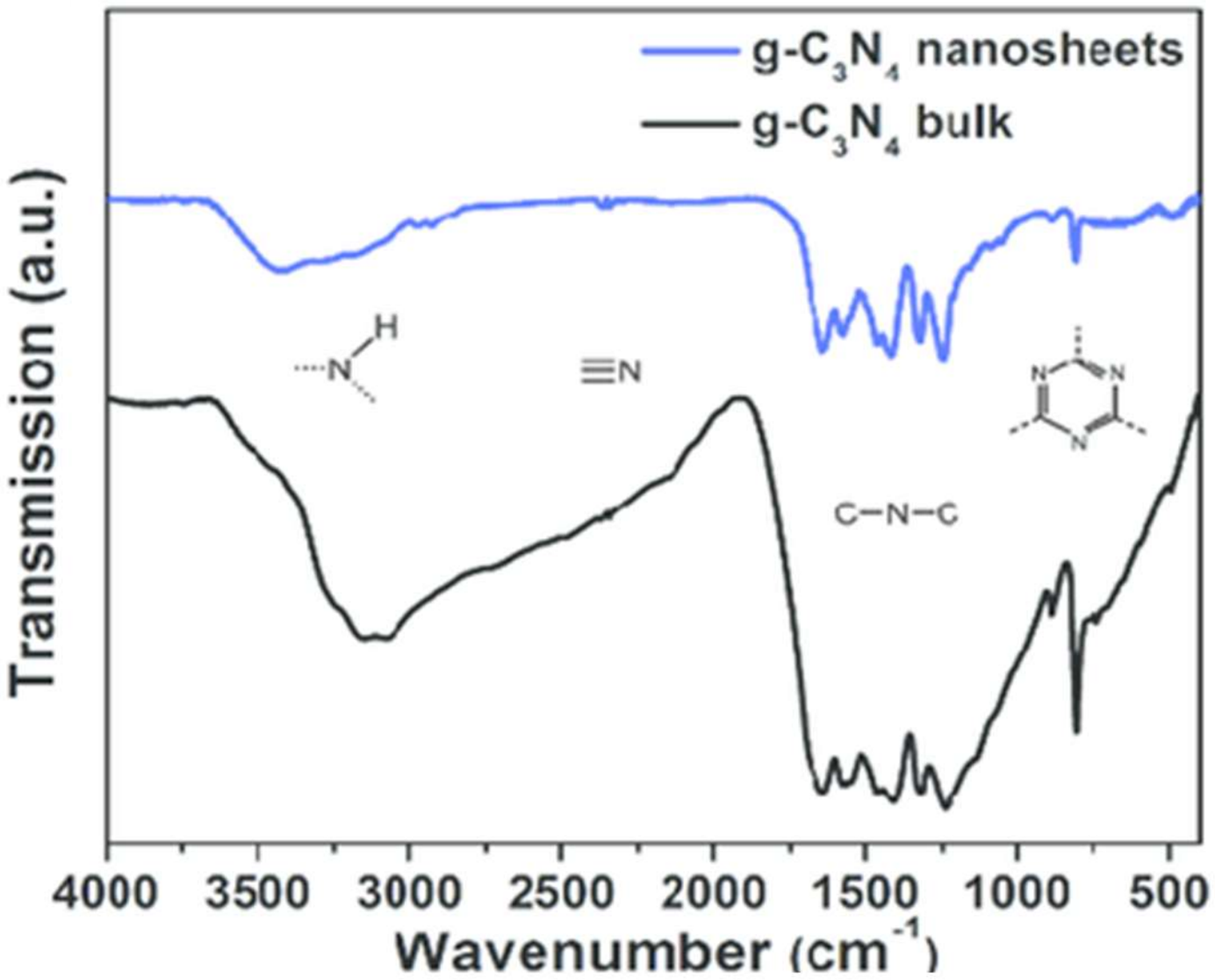

**Fig. 13** The FTIR spectra of the bulk and nano-sheets along with the major components for the corresponding peaks **[61].**

**X-Ray Photoelectron Spectroscopy** of materials reveal the presence of variety of atoms in terms of survey scan spectra and relative percentage of hybridization states in terms of core level spectra. XPS spectra represents the number of detected electrons as a function of the binding energy of the detected electrons within a material when it is exposed to a beam of X-rays. It is a surface sensitive technique and can only measure a material surface within 10 nm. The XPS spectrum of g-$C_3N_4$ gives information about the chemical states of the material. The C*1s* spectrum shows two peaks at about 284.6 eV and 288.0 eV. The 288.0 eV peak is further deconvoluted into two Gaussian-Lorenzian peaks. The main contribution peak at 287.9 eV is attributed to the *$sp^2$* hybridized carbon bonded to N inside the triazine rings, while the peak at 288.6 is assigned to the *$sp^2$* -hybridized carbon in the triazine ring bonded to the $–NH_2$ group. The peak at 284.6 eV is typically ascribed to *$sp^2$* C=C bonds **[33, 62].** The spectrum from N*1s* was deconvoluted by three components at binding energies of 399.2, 401.2 and 404.4 eV. The main peak at 399.2 eV is assigned to C-N-C coordination, while the peak at 401.2 eV is assigned to *$sp^3$* N-$(C)_3$ of the tertiary N from the bridges and the center of the tri-striazine units. The weak shoulder component at 404.4 eV is due to $\pi \rightarrow \pi^*$ excitations **[63]**. *Yang et al.* observed that the lowest energy contribution $C_3$ at 283.8 eV which is typically assigned to graphitic C=C or/and the cyano- group in the literature becomes clearer in the case of nanosheets with respect to that of bulk g-$C_3N_4$, indicating little nitrogen is removed during the sonication process **[64].** Sometimes, weak O signals are also observed in the XPS spectra which can be ascribed to the absorbed oxygen species on sample surface and the slightly higher O peak for nanosheets may be due to their high specific surface area and protonation by $H_2SO_4$ **[44].** Deconvolution of the XPS signals also reveals a weak additional signal at 401.4 eV, indicative of amino functions carrying hydrogen (C–N–H) which is attributed to the charging effects **[8].** For g-$C_3N_4$ nanosheets, a new peak located at 403.2 eV becomes noticeable which is due to the protonation of g-$C_3N_4$ nanosheets, which causes CN heterocycles and cyano groups to be positively charged **[53].** According to the percentage of C and N determined by the XPS, *Niu et al.* observed that the surface atomic ratio of C to N increased from 0.65 for bulk g-$C_3N_4$ to 0.83 for nanosheet due to the oxidation of nitrogen **[34].** Also, while shifting from nanoplates to nanorods *Bai et* *al.* found that although the XPS peaks were almost same for nanoplates and nanorods, nanorods had a higher intensity of the peak around 398.5 eV (C-N=C) and 400.1 eV (N-(C)$_3$) suggesting that the nanorods had more heptazine units and less surface defects **[43].**

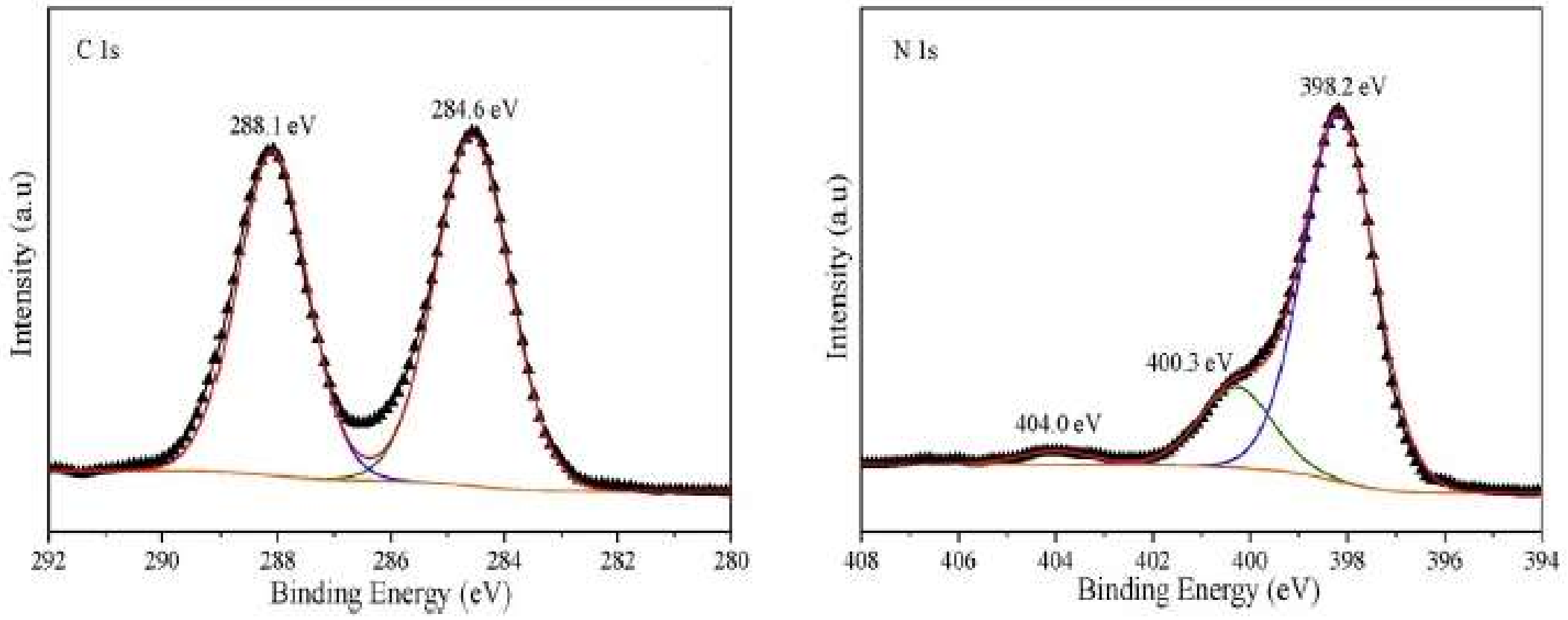

**Fig. 14** XPS spectra of bulk g-$C_3N_4$ (a) for the carbon core and (b) for the Nitrogen core. **[39]**.

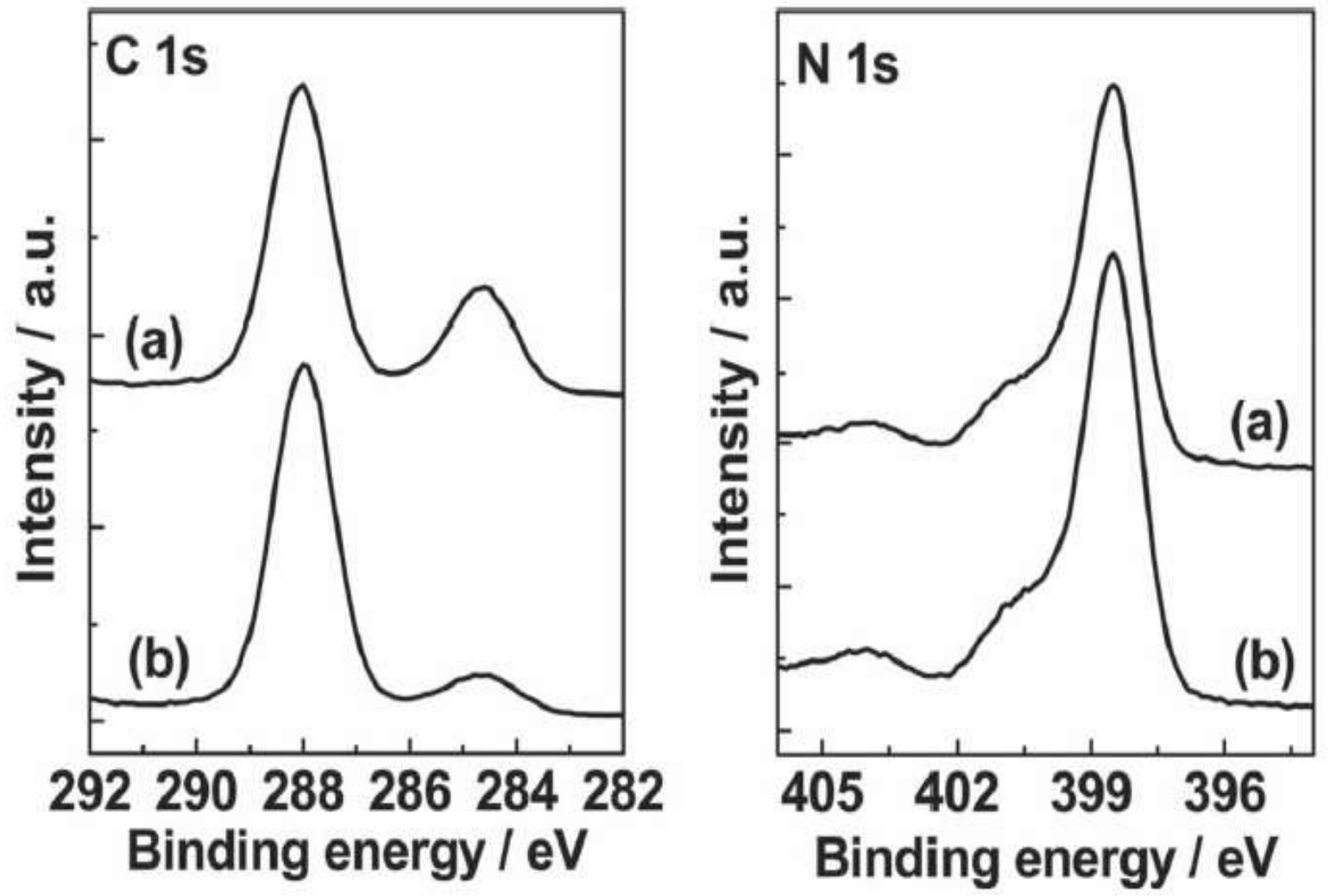

**Fig. 15** C*1s* and N*1s* XPS spectra of: a) bulk g-$C_3N_4$, and b) g-$C_3N_4$ nanosheets. The peak of C*1s* at 284.6 eV arises from the adventitious carbon in the samples. **[34].**

## Optical Properties

**UV-Vis absorption spectra:** UV-Visible Diffuse Reflectance Spectroscopy (DRS) is an important method to explore the band-gap of the solid materials. The DRS of bulk g-$C_3N_4$ shows a single light absorption edge at around 450 nm. This absorption peak is characteristic of g-$C_3N_4$ polymers and corresponds to the intrinsic $\pi \rightarrow \pi^*$ electronic transition **[65-66].** This indicates that the material can be excited under visible light irradiation **[67].** The spectra, however, depends on the heating temperature as well as the precursor used. *Zou et al.* reported the absorption edge at 443 nm for bulk g-$C_3N_4$ prepared by direct pyrolysis of guanidine hydrochloride **[60].** *Yan et al.* compared the DRS spectra of bulk g-$C_3N_4$ prepared by heating Cyanamide at 550 ºC with the DRS of bulk g-$C_3N_4$ prepared by heating melamine at different temperatures. They observed that the absorption edges of the samples obtained from heating melamine shifts remarkably to larger wavelengths with increasing heating temperature. The band gap of the samples also decreased from 2.8 eV to 2.75 eV when heating temperature is increased from 500 to 580 °C **[42].** The red shift of the absorption spectra of bulk g-$C_3N_4$ is due to the extended delocalization of 2D electrons**. [65, 68]** *Jia et al.* reported that the bandgap of g-$C_3N_4$ prepared by melamine and urea is larger than the g-$C_3N_4$ prepared using only melamine (2.73 eV) but smaller than g-$C_3N_4$ prepared using urea (2.87eV) **[40].** As the dimensions are decreased, the spectra blue shifts. Contrary to the bulk materials, this blue shift goes on increasing with increasing temperature. *Tahir et al.* observed that for nanofibres the band edge is shifted from 465 nm to 443 nm and the band gap is changed from 2.67 eV to 2.80 eV. This is due to more efficient packing, electron coupling as well as the quantum confinement effect. This also proves that the nanofibres have lesser defects and a slower recombination rate **[29]**. The absorption spectra of g-$C_3N_4$ nanosheets do extend to the whole visible light region and even to the IR region enhancing the absorbance of light. For a monolayer nanosheet, *Xu et al.* reported blue shift in absorption edge from 470 nm for bulk to 425 nm corresponding to the increase in bandgap from 2.64 eV to 2.92 eV **[33].** Again, this is attributed to the strong quantum confinement effect and decrease in conjugation length.

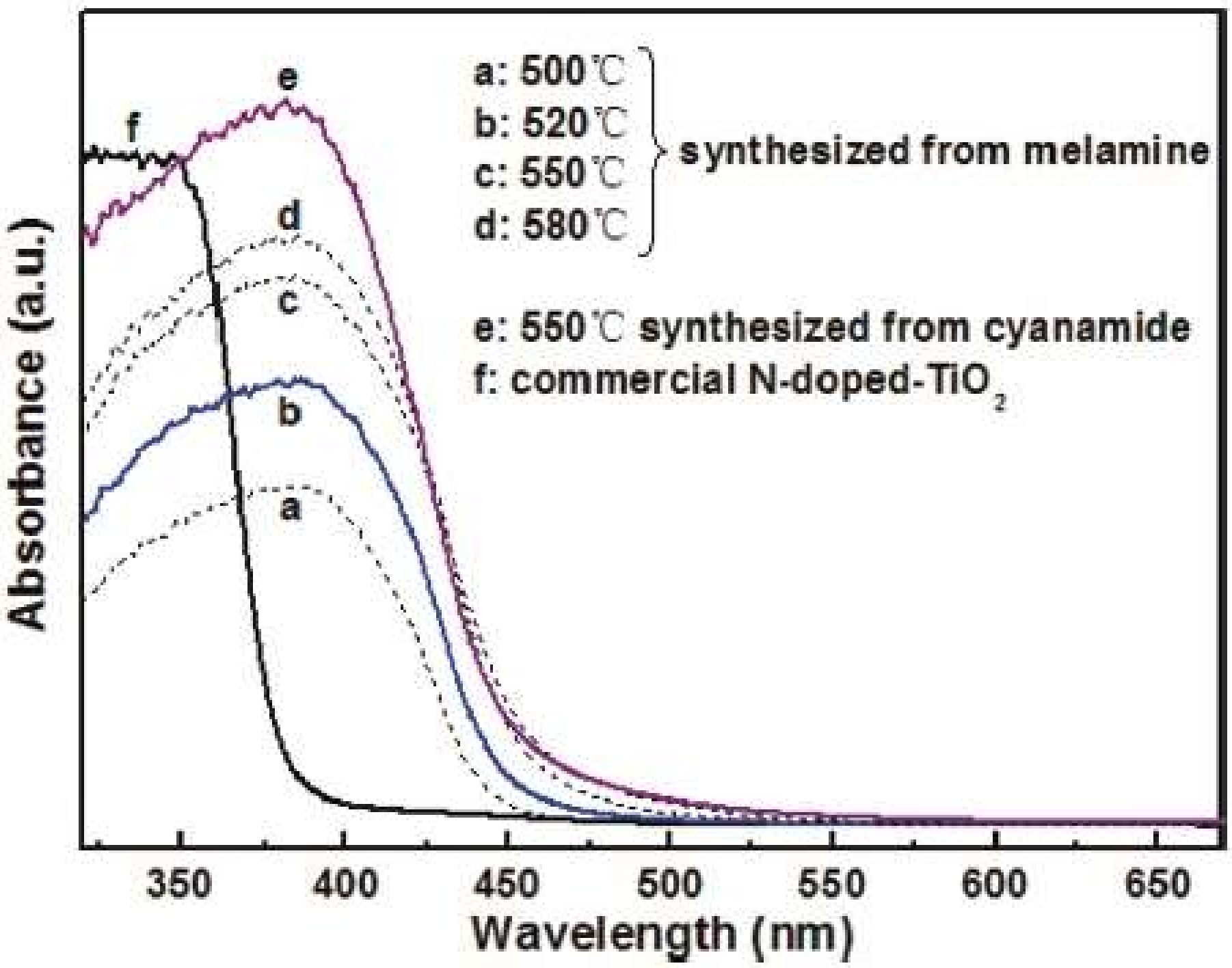

**Fig.16** UV-vis absorption spectra for commercial N-doped-$TiO_2$ and g-$C_3N_4$ samples obtained by heating the different precursors at different temperatures **[42].**

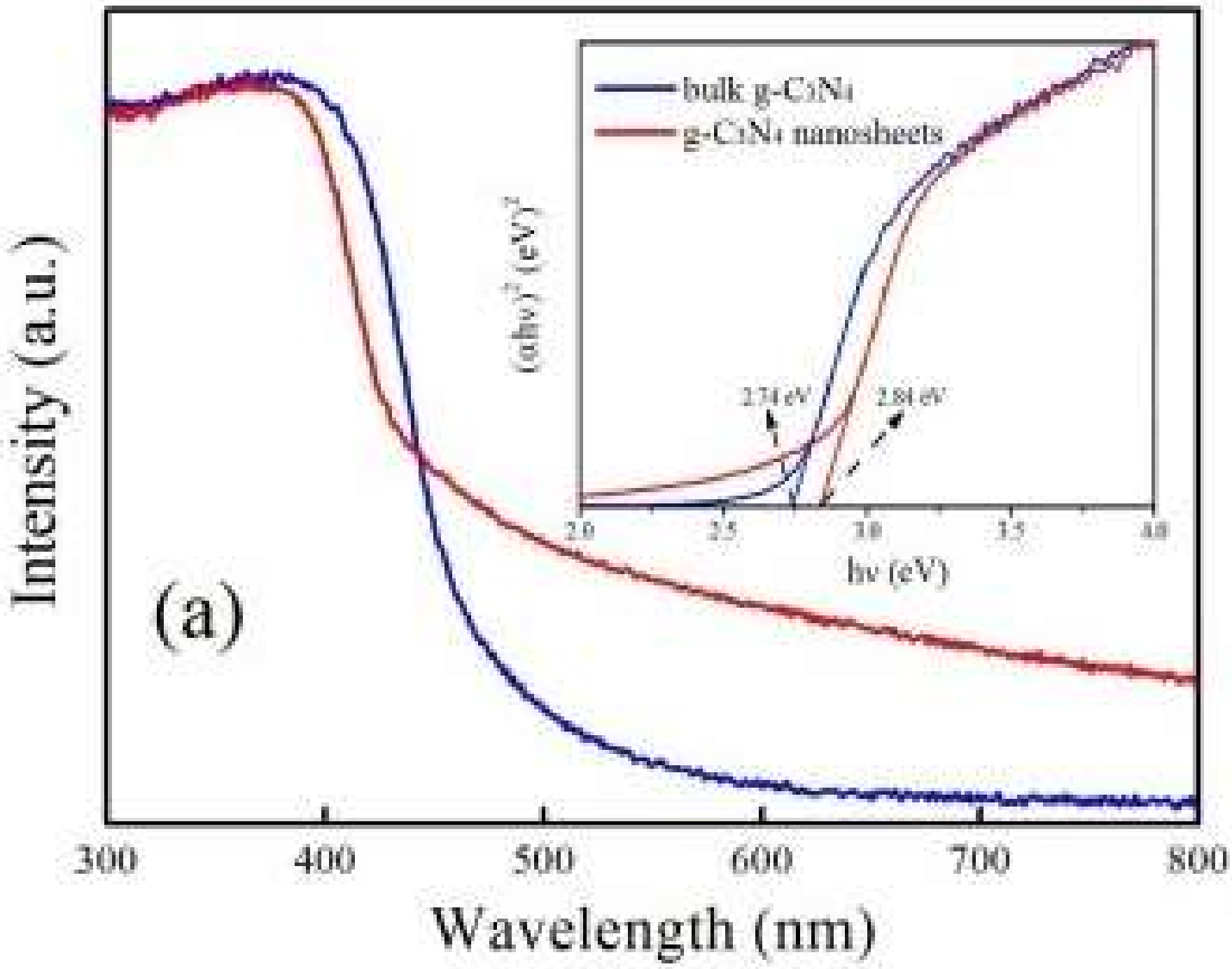

**Fig. 17** Shows the difference in the UV–vis DRS spectra i.e band gaps (inset) of bulk g-$C_3N_4$ and its nanosheets **[60].**

**Photoluminescence (PL):**

The PL emission for bulk g-$C_3N_4$ shows a strong peak at 464 nm. *Zhang et al.* studied the changes in the emission peak along with the change in temperature (300 ºC to 650 ºC) for bulk g-$C_3N_4$ and observed that the PL of the product slightly red shifts with the processing temperature. Under 280 nm wavelength excitation, the center of the peak shifts from 400 to 510 nm with increasing temperature. This PL excitation band can be attributed to the transition between the lone pair valence band and the π* conduction band **[69].** This red shift can be explained by the extension of the g-$C_3N_4$ network at higher temperatures. When more heptazine is connected by the amino group, the π states will hybridize into a broad state, causing the bandgap narrowing of the *sp²* C–N clusters **[70-71].** Zhang et al. studied the PL behavior for nanosheets and noted that as we move on to smaller dimensions, the PL spectra blue shifts due to the quantum confinement effect with the conduction band and valence band shifting to the opposite directions. They also reported the pH dependence of the PL of the nanosheets and attributed this dependence to the presence of free zigzag sites. They also reported the PL quantum yield to be 19.6 % which was much higher than that of bulk g-$C_3N_4$ (4.8 %) **[31].** *Wang et al.* reported that for nanotube type graphitic carbon nitride (GCN) a broad PL excitation band with peak at 460 nm was observed **[30].** In case of nanofibres, *Tahir et al.* reported that the graphitic carbon nitride nanofiber (GCNNF) had larger bandgap and the PL peak had a lower intensity than that of bulk graphitic carbon nitride, due to the presence of lesser defects in the nanofibres and a slower recombination rate of electrons and holes **[29].**

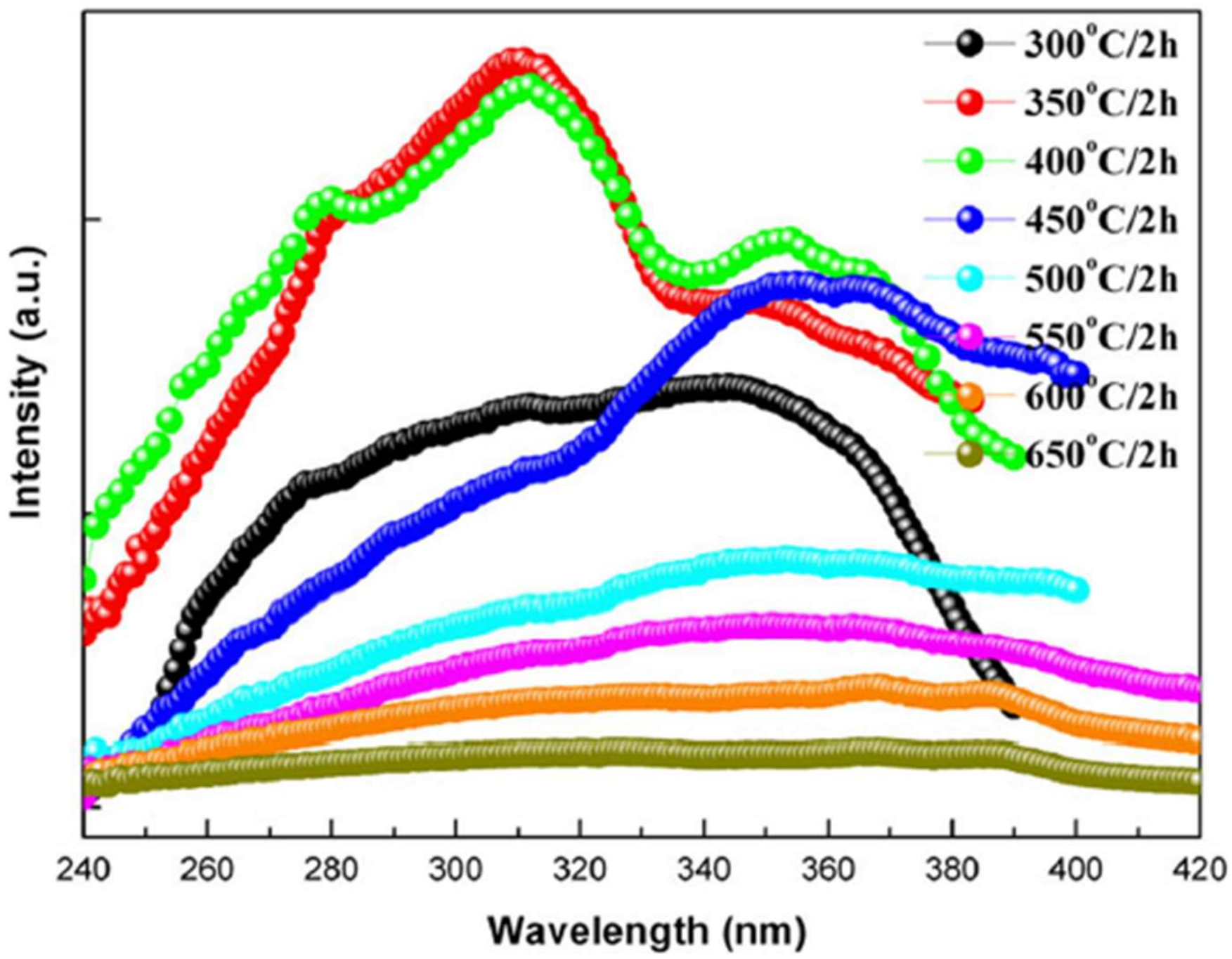

**Fig. 18** The PL excitation spectra of the carbon nitride products synthesized via the thermal condensation of melamine at different temperatures for 2 h indicating variation in the optical band-gap **[69].**

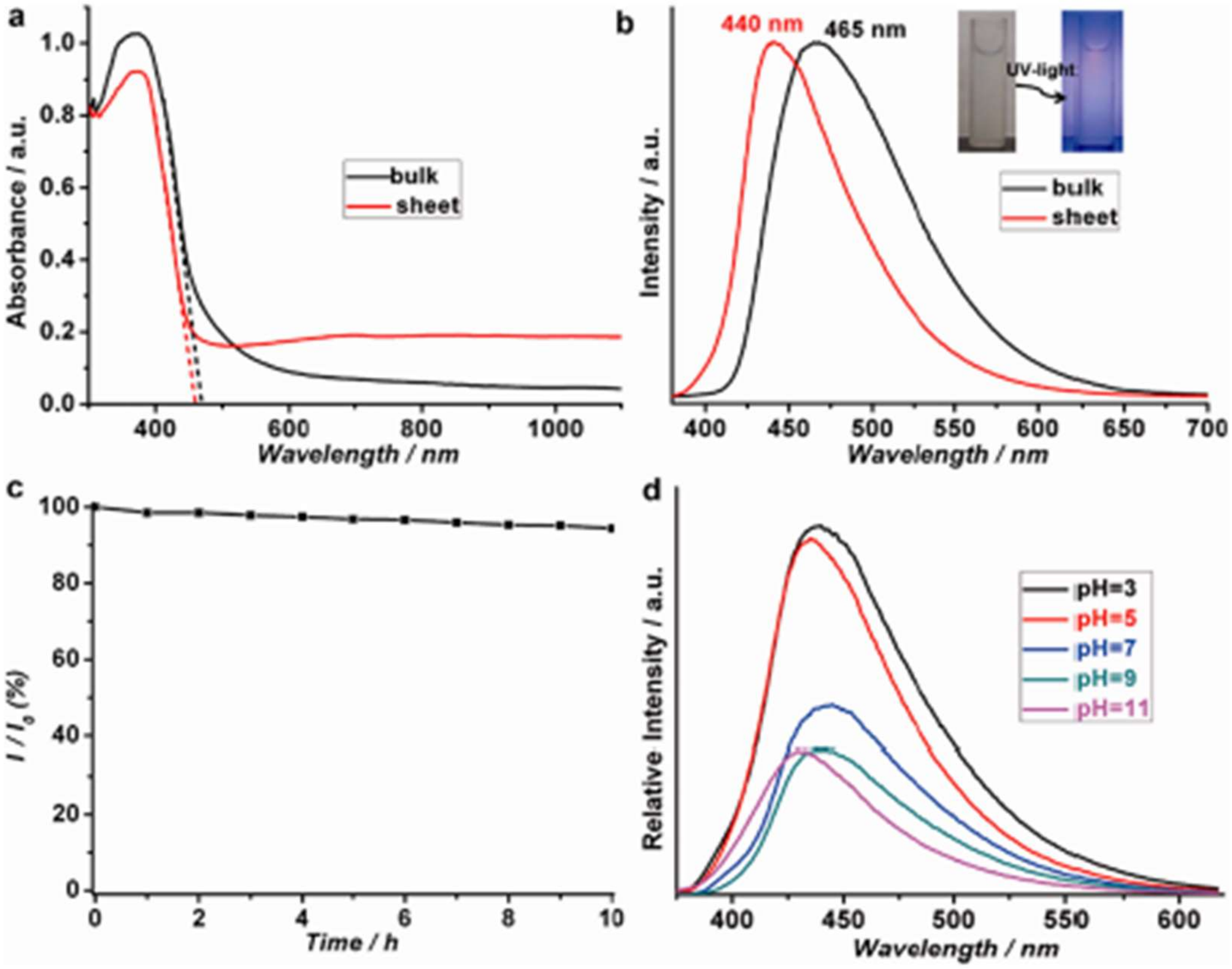

**Fig. 19 (a)** UV−visible absorption spectra. (b) Normalized photoluminescence spectra of bulk g-$C_3N_4$ and ultrathin g-$C_3N_4$ nanosheets; inset of (b) is the color change of ultrathin g-$C_3N_4$ nanosheets solution before and under UV light illumination. (c) Dependence of PL intensity on UV-light illumination time for ultrathin g-$C_3N_4$ nanosheet solution. (d) The pH-dependent PL behavior of ultrathin g-$C_3N_4$ nanosheet aqueous solution **[31].**

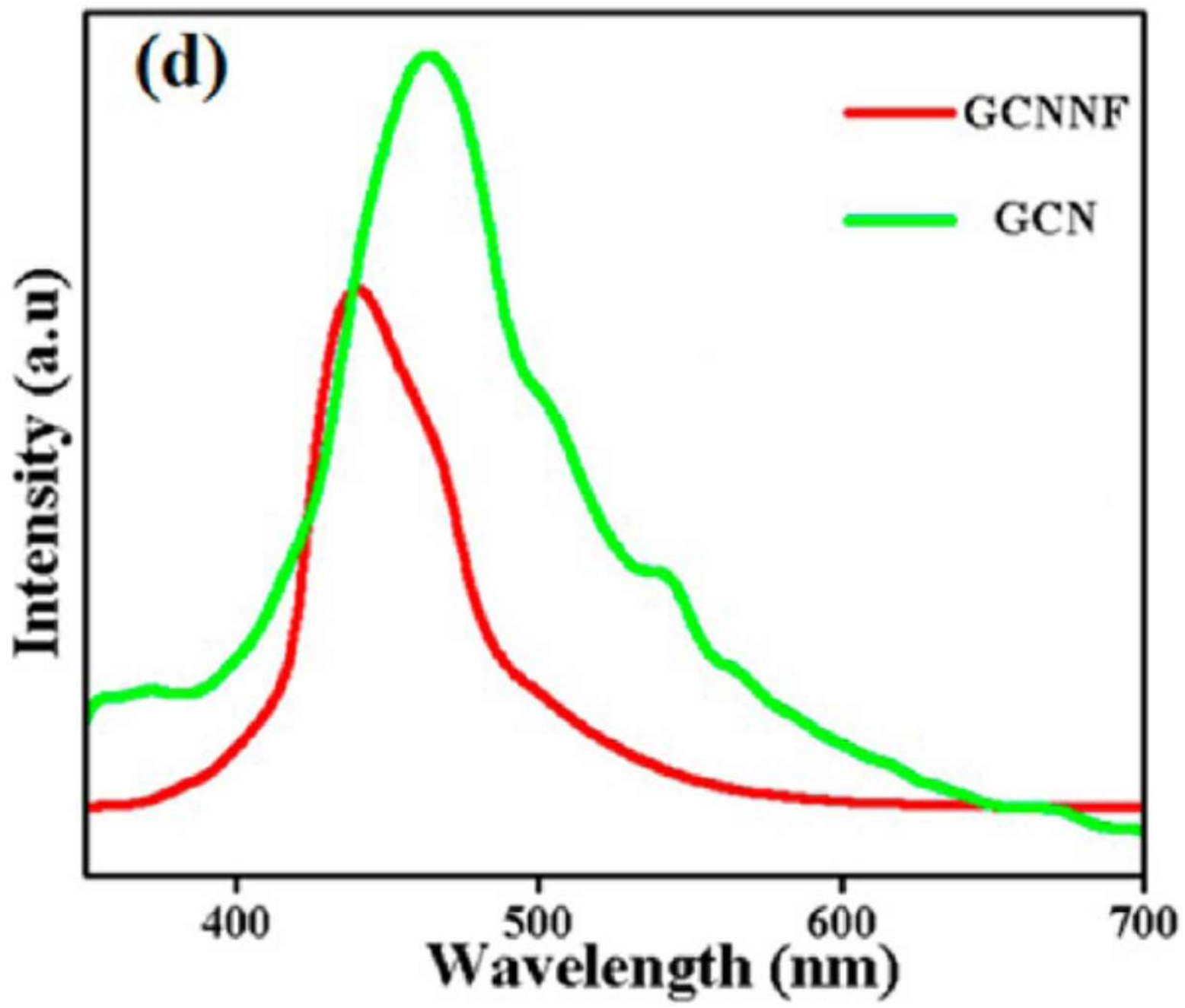

**Fig. 20 (b)** PL spectrum of GCNNF and GCN. It shows that the graphitic carbon nitride nanofiber (GCNNF) had larger bandgap and the PL peak had a lower intensity than that of bulk graphitic carbon nitride, due to the presence of lesser defects in the nanofibres and a slower recombination rate of electrons and holes **[29].**

**Thermal Properties**

g-$C_3N_4$ is non-volatile up to 600°C. *Li et al.* analyzed the effect of temperature on g-$C_3N_4$ formation using melamine as a precursor **[72]**. Thermo-gravimetric analysis (TGA) and differential thermal analysis (DTA) in the presence of air shows a sharp weight loss step in the temperature range of about 550 to 700 °C as seen in Fig. 21, which is attributed to the oxidation of the g-$C_3N_4$ to form graphite and $N_2$. The exothermic peak at 740°C may be caused by the oxidation of graphite to form $CO_2$ **[72].** An initial weight loss is also observed below 200 °C, which arises from the evaporation of adsorbed water to the sample surface. *Yan et al.* on doing thermal characterization of g-$C_3N_4$ (synthesized via heating melamine) through *DSC-TG*A observed that the strongest endothermal peak was at temperatures 297-390 °C where the weight of the sample decreased rapidly as visible in Fig.22. Two weak endothermal peaks were also found at 545-630 °C, whereas two exothermic reaction peaks attributing to the disappearance of material were observed at 660-750 °C. They also observed that $g\text{-}C_3N_4$ was more stable in a semi closed ammonia atmosphere than in an open system **[42].**

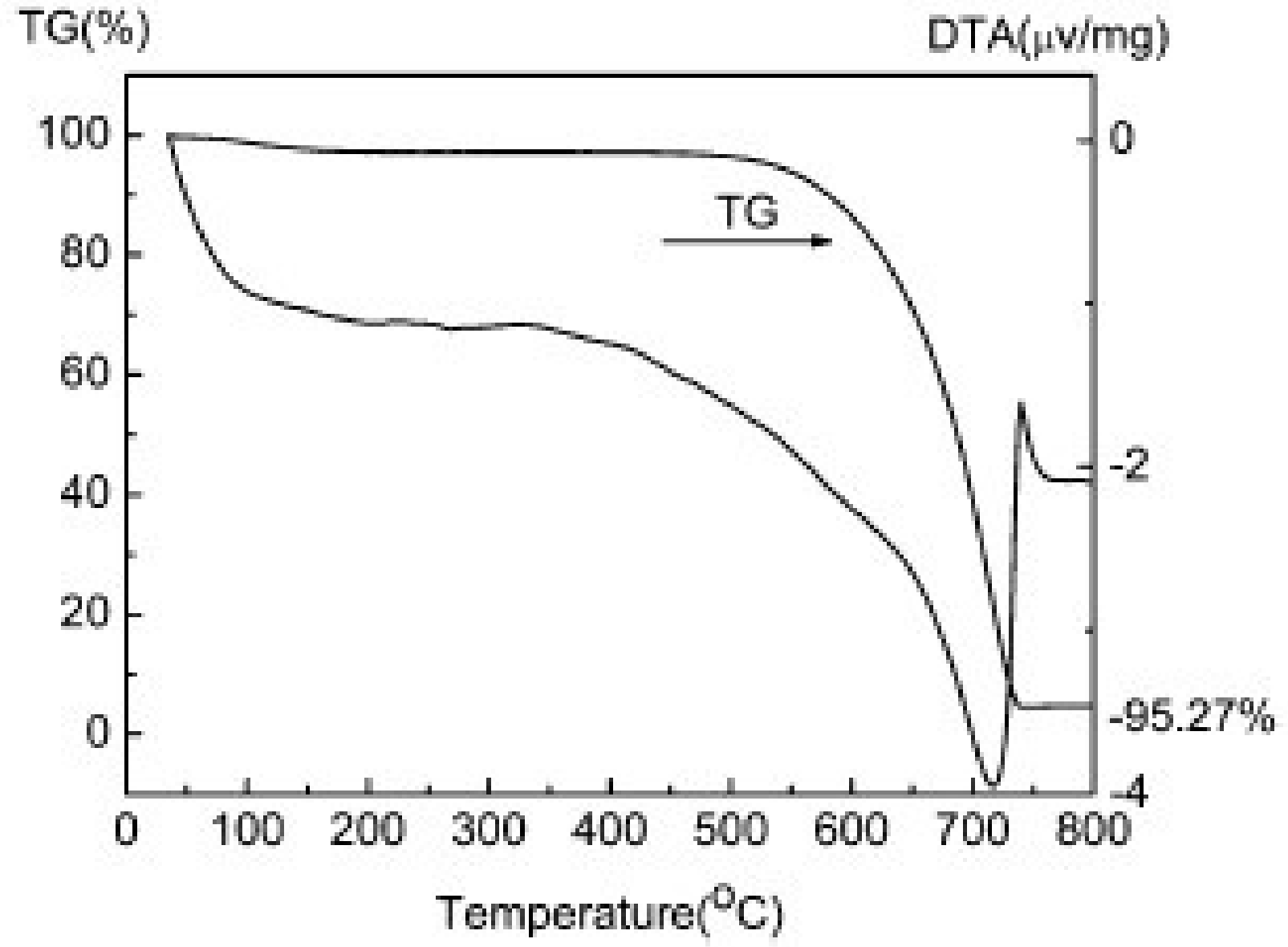

**Fig. 21** The TGA-DTA spectrum of g-$C_3N_4$. It shows stability of g-$C_3N_4$ upto 550 °C. **[72].**

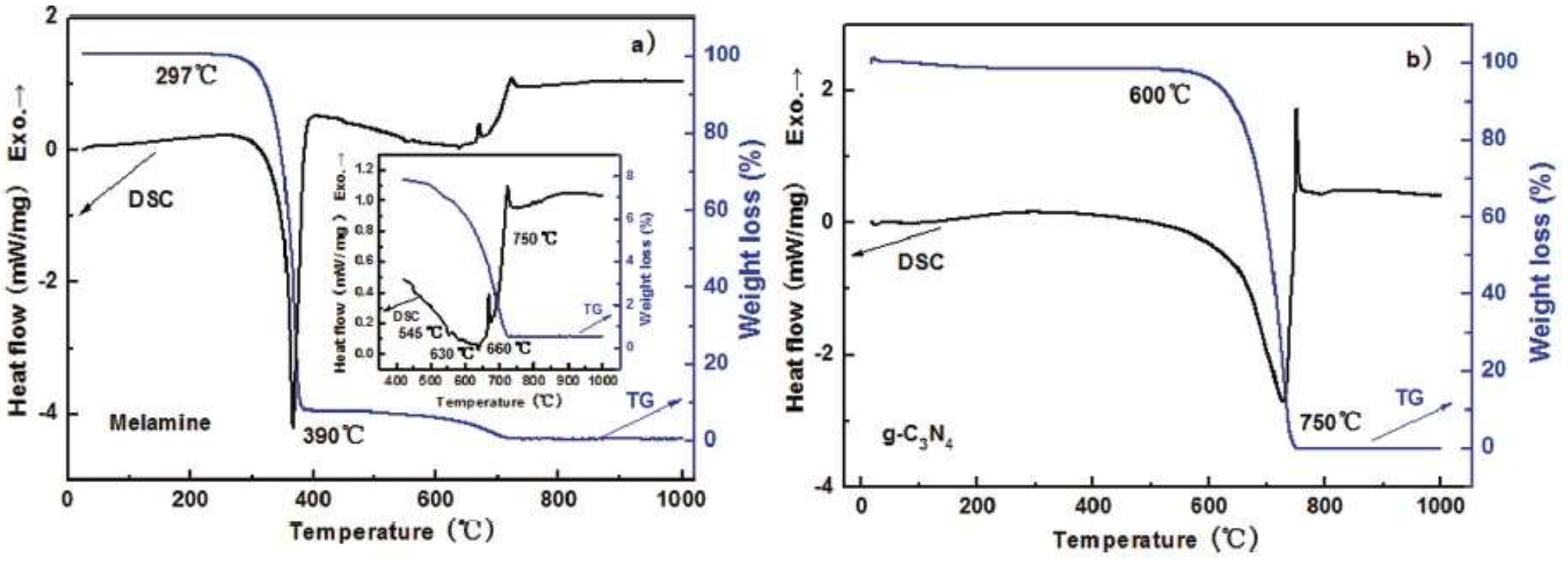

Fig. 22 TGA-DSC thermograms for heating (a) the melamine and (b) the $g\text{-}C_3N_4$ obtained by heating melamine at 520 °C **[42].**

## Applications

Since $g\text{-}C_3N_4$ is a visible light polymeric i.e nonmetallic, highly stable n-type semiconductor having a band gap of ~2.7 eV (medium bang-gap) corresponding to an optical wavelength of ~ 450 nm, it finds scope for wider range of applications like photodetectors, photocatalytic $H_2$ and $O_2$ generation, photocatalytic degradation of dyes and pollutants sensors, super capacitors, Li-ion batteries, bio-imaging and anticorrosive coating etc **[10, 73]**. Each of these applications and the performances achieved using $g\text{-}C_3N_4$ has been summarized next.

**$g-C_3N_4$ as Photo-catalysts**

*Wang et.al* in 2009 was the first one to use $g-C_3N_4$ as a photo-catalyst for water splitting. They observed that the band gap of $g-C_3N_4$ was large enough to overcome the endothermic character of water splitting (1.23 eV theoretically) **[74].** Its band gap also falls in the range of visible light absorption (< 3.1 eV) **[75].** In spite of these unique properties, it has been challenging to use bare $g-C_3N_4$ as a photo-catalyst due to several drawbacks such as low quantum efficiency (0.1 %) at 420-460 nm **[76-78]** high electron-hole recombination rate, low surface area, and smaller active sites **[79].** It has strong reduction reaction properties owing to the high potential of the conduction band but inferior oxidation capabilities due to its valence band located at about 1.4 eV vs. NHE (Normal Hydrogen Electrode) resulting in a small thermodynamic driving force for water or organic pollutants oxidation **[80].** Several approaches have been taken to overcome these issues to modify the structure of $g-C_3N_4$ by doping it with impurities or forming its hetero-structures.

In the properties section, how the efficiency of $g-C_3N_4$ varies as we shift to the lower dimensions has been discussed. The PL quantum yield of $g-C_3N_4$ nanosheets increase from ~ 4.8 % to 19.6 % **[31].** In case of nanorods, the band gap decreases from 2.75 eV to 2.66 eV depicting a lower recombination rate of electrons and holes under visible light irradiation **[43]**. *Niu et al.* detected the amount of $OH^-$ radicals generated using nanosheets and found that the $H^-$ evolution rate of $g-C_3N_4$ nanosheets under UV-Vis light was 170.5 $\mu molh^{-1}$, which is 5.4 times higher than its bulk counterpart. The surface area is increased from 50 $m^2g^{-1}$ to 306 $m^2g^{-1}$ **[34]**. Given below is a table of plain g-$C_3N_4$ counterparts and their respective hydrogen evolution rate.

**Table: 1** Summary of use of g-$C_3N_4$ as photo-catalyst.

| Photocatalyst | Hydrogen evolution ($\mu molh^{-1}$) | Amount of photo-catalyst (mg) | Light source | AQE | Reference |
|---|---|---|---|---|---|
| Pt-g-C3N4/CdS With 2% Pt co-catalyst | 35,300 | 20 mg | 300 W Xe lamp (λ > 420 nm) | 24.8% at 420 nm | (Nagella et al., 2023) **[79]** |
| Zn doped g-C3N4 With 1% Pt co-catalyst | 78.7 | 75mg | 250W mercury lamp (λ≥420nm) visible light | | (Fuentez-Torres et al., 2021) **[81]** |
| g-C3N4 (ZIS-S/CN) With 3% Pt co-catalyst | 6100 | 50 mg | 300 W Xe lamp (λ≥420nm) visible light | 12.9% at 400 nm | (Qin et al., 2020) **[82]** |
| g-C3N4.7 with 3% Pt | 4910 | 50 mg | 250 W Xe lamp (λ ≥ 190 nm) visible light | 14.07 % at 479 nm | (Antil et al., 2021) **[83]** |
| g-C3N4/saPt | 50 mg | 14,740 | 300WXe lamp (λ ≥ 400 nm) visible light | 38.8% at 435 nm | (L. Zhang et al., 2020) **[84]** |
| G-g-C3N4 | 18,900 | 50 mg | 350 W Xe lamp | 19.4% at 420 nm | (Yao et al., 2021) **[85]** |
| g-C3N4/ PTEtOH With 1% Pt co-catalyst | 2424 | 100 mg | 300 W Xe lamp (λ ≥ 420 nm) visible light | 9.5% at 405 nm | (Zhao et al., 2021) **[86]** |
| g-C3N4/ In2O3 cube With 3% Pt co-catalyst | 1917 | 20 mg | 300 W Xe lamp (λ ≥ 400 nm) visible light | 7.73% at 420 nm | (W. Wang et al., 2021) **[87]** |
| g-C3N4/ SnO2 With 3% Pt as co-catalyst. | 2544 | 10 mg | 300 W Xe lamp (λ > 400 nm) | 9.63% at 420 nm | (Yan et al., 2021) **[88]** |
| $MnWO_4$/ g-$C_3N_4$ With 0.5% Pt co-catalyst | 2820.2 | 50 mg | 200 W Xe lamp (λ ≥ 420 nm) visible light | 16.51 % at 400 nm | (Zhou et al., 2022) **[89]** |
| $Bi_3TaO_7$ / OR- g-$C_3N_4$ With 3% Pt co-catalyst | 4891 | 25 mg | 300 W Xe lamp (λ ≥ 420 nm) visible light | 4.1% at 420 nm | (Shi et al., 2022) **[90]** |
| $SnNb_2O_6$/g-$C_3N_4$ | 1427.12 | 20 mg | 300 W Xe lamp (λ ≥ 420 nm) visible light | 3.95% , at 365 nm | (H. Wang et al., 2022) **[91]** |
| pg-$C_3N_4$/$Cd_{0.5}Zn_{0.5}S$diethylenetriamine | 12627 | 20 mg | 300 W Xe lamp (λ ≥ 420 nm) visible light | 37.7% at 420 nm | (Chen et al., 2021) **[92]** |
| Ag/CNU-CNT 0.3% Pt co-catalyst | 10100 | 10 mg | 500 W metal halide lamp | 39% at 420 nm | (Gogoi et al., 2021) **[93]** |
| g-$C_3N_4$/ $Bi_4Ti_3O_{12}$/ | 56200 | 45 mg | 500 W Xe | 23.1% | (Kumar |
| Bi4O5I2 | | | lamp (λ > 420 nm) | at 420 nm | et al., 2021) **[94]** |
| g-C3N4/ W18O49/CdS With 1% Pt co-catalyst | 11658 | 100 mg | 300 W Xe lamp (λ > 420 nm) | 26.73 % at 420 nm | (Yang et al., 2021) **[95]** |
| Pt-g-C3N4/CdS With 2% Pt co-catalyst | 35,300 | 20 mg | 300 W Xe lamp (λ > 420 nm) | 24.8% at 420 nm | (J. Wang et al., 2021) **[96]** |
| g-C3N4/Co2P | 4752 | 50 mg | 300 W Xe lamp (λ ≥ 420 nm) visible light | 11.2% at 420 nm | (J. Zhang et al., 2021) **[97]** |
| g-C3N4/CdS | 6.38 | 10 mg | 500 W metal halide lamp | 29.6 % | (Rambabu & Peela, 2023) **[98]** |

**Enhancing electrochemical properties by Heteroatom Doping**

Heteroatom doping has been an efficient method for the enhancement of defects and thus performance enhancement for a long time. The same is the case with g-$C_3N_4$. *Ma et al. using* first-principles studies showed that doping of g-$C_3N_4$ with nonmetals such as Phosphorus or Sulfur modifies its electrochemical properties, where S atoms substitute for the edge of N atoms, whereas P atoms preferentially situate in the interstitial sites of planar g-$C_3N_4$ **[99-100].** Oxygen doping has also been found to improve the photocatalytic activity. The enhanced photocatalytic activity is due to the more adsorption sites and more active sites, the enhanced redox ability and improved electron transport ability, which leads to less recombination and a more efficient separation of photogenerated electron and hole pairs **[13, 101].** In all of these cases, it has been observed that the visible light absorption range has increased. *Luo et al.* observed that the incorporation of HCl in g-$C_3N_4$ can switch its semi-conductivity from n-type to p-type due to the acceptor level formed above the valence band by the $Cl^-$ atoms **[102].** *Lan et al.* synthesized Br modified g-$C_3N_4$ via condensation of urea with $NH_4Br$ and observed enhanced optical absorption and photocatalytic activities without the destruction of major architecture of g-$C_3N_4$ polymer **[103].** *Zhu et al.* chemically activated g-$C_3N_4$ and found that while the functional groups and crystal structure remained intact, the light absorption range was increased due to the introduction of O and H elements into the C-N framework **[104].**

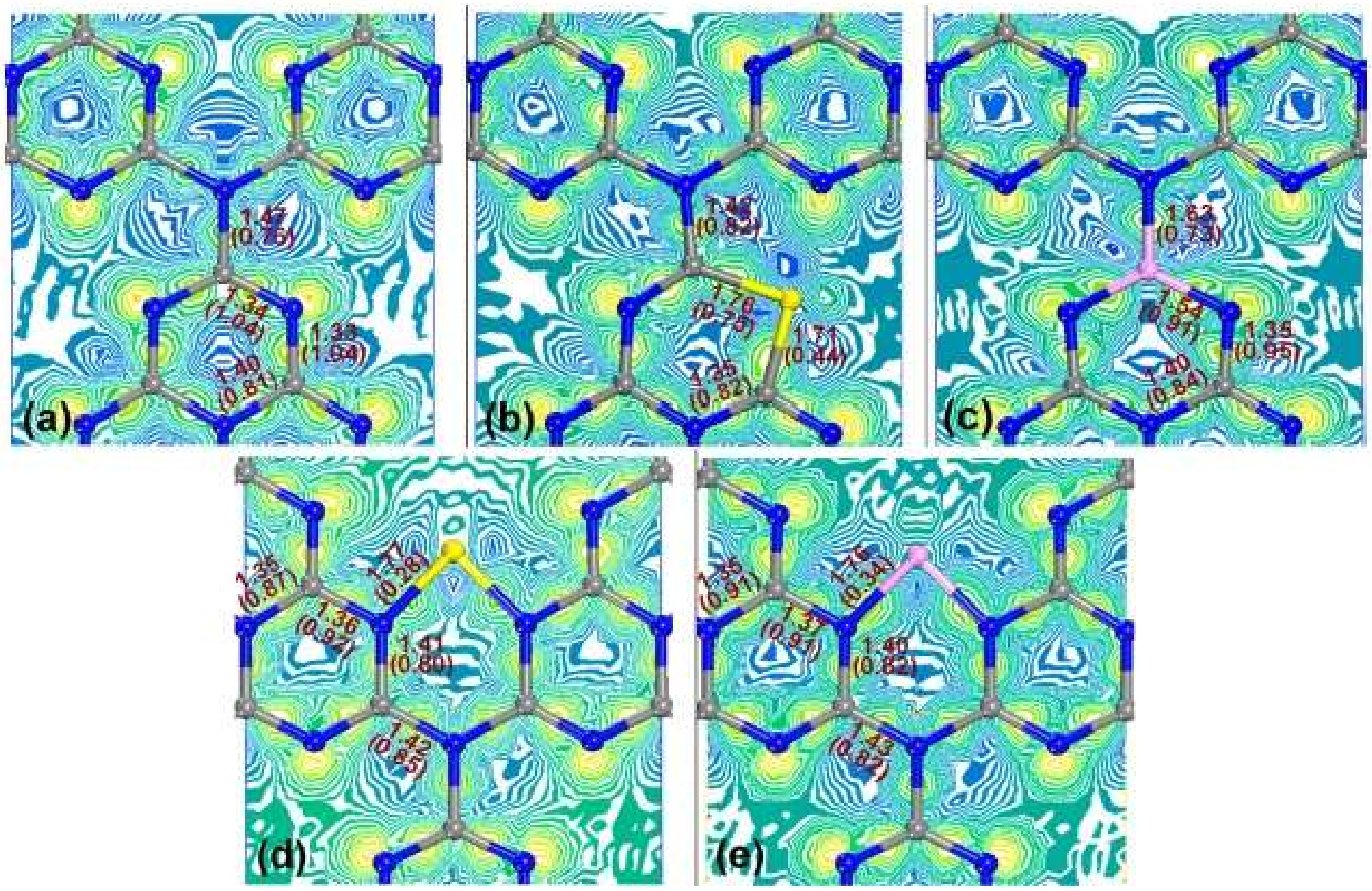

**Fig. 23** Shows that difference in charge density contour maps in planar and part of bond length and corresponding bond overlap population of g-$C_3N_4$ systems: (a) pure, (b) SN1, (c) PC1, (d) Si, (e) Pi. Gray and blue spheres represent the C and N atoms, respectively. In addition, yellow and purple spheres represent the S and P atoms (impurities), respectively. It was observed that doping of g-$C_3N_4$ with nonmetals such as Phosphorus or Sulfur modifies its electrochemical properties **[105-106].**

**Defect engineering to enhance the photocatalytic properties**

*Dong et al.* studied carbon vacancy modified g-$C_3N_4$ nanosheets and observed that the enhanced efficiency in the photocatalytic NO reduction is due to the carbon vacancies serving as traps for the photogenerated electrons and providing preferential adsorption sites for NO **[107].** Whereas *Li et al.* observed that the nitrogen-deficient g-$C_3N_4$ has an increased number of active sites for the photocatalytic reaction and it facilitates the efficient separation of photo-generated electrons and holes **[108].** *Niu et al.* observed that the nitrogen deficit g-$C_3N_4$ synthesized via increasing the poly-condensation temperature of dicyanamide showed a slightly narrowed bandgap and extended visible light absorbance between 450 and 600 nm due to the abundant nitrogen vacancy-related $C^{3+}$ states in the band gap **[109].** *Ruan et al.* synthesized g-$C_3N_4$ nanosheets using melamine and urea. They observed that the introduction of urea enhanced the surface area as well as the defects of nitrogen vacancies, which acted as active trap sites in g-$C_3N_4$, increasing the apparent quantum efficiency (AQE). Photocatalytic materials harness solar energy to generate hydrogen ($H_2$) through the water-splitting process. The efficiency of this photon-to-hydrogen conversion is typically evaluated using the AQE. AQE quantifies the proportion of incident photons that are effectively utilized by the photocatalyst to produce hydrogen. It is calculated based on the measured rate of hydrogen evolution and the flux of incoming photons, using the following formula: AQE (%) = (2 × rate of $H_2$ evolution) / rate of incident photons for $H_2$ evolution to 74% **[110].**

## Heterojunction /Composite Formation

Various composites of g-$C_3N_4$ have been synthesized and studied in order to enhance their photocatalytic activities and optical absorption range. All of them show greater efficiency due to the formation of different types of heterojunctions. Firstly, the composites were made with the previously used photo-catalysts such as ZnO, $TiO_2$ etc. $TiO_2$ having a larger UV response range was thought to enhance the charge photogeneration in g-$C_3N_4$ **[111].** It was observed that the light absorption range was broadened and was from 300-450 nm for the composite. The type II heterojunction formation led to lesser electron hole recombination and greater charge separation efficiency. A type II heterojunction is also referred as staggered band alignment heterojunction, is a junction formed between two semiconductors where the conduction band minimum and valence band maximum of one material are both at higher energy levels than those of the other. This configuration causes photo-generated electrons and holes to separate spatially into different materials, enhancing charge separation and reducing recombination, making it highly beneficial for applications like photocatalysis and solar energy conversion **[11, 112].** Similar case happens on combining g-$C_3N_4$ with ZnO resulting into formation of a type II heterojunction **[113]**. *Liu et al.* studied the efficiency of ZnO/g-$C_3N_4$ composite via the photo-oxidation of RhB and photo-reduction of $Cr^{6+}$, they found that for 5.0 wt % g-$C_3N_4$, the photooxidation rate constant for RhB was 0.0367 $min^{-1}$, while it was only 0.0112 $min^{-1}$ for pure $C_3N_4$ and the photoreduction rate constant of $Cr^{6+}$ was more than five times that observed for pure $C_3N_4$ **[114].** $MoS_2$/g-$C_3N_4$ is also an efficient heterojunction for photodegradation activities **[115, 116]** *Wang et al.* synthesized $MoS_2$/g-$C_3N_4$/GO with 5.6 wt % $MoS_2$/g-$C_3N_4$ and found the hydrogen evolution rate was increased to 1.65 $mmolh^{-1}g^{-1}$ from 1.06 $mmolh^{-1}g^{-1}$ for $MoS_2$/g-$C_3N_4$ indicating that the GO layer helps in efficient collection of electrons in $MoS_2$ and holes in g-$C_3N_4$ **[117].** $MoO_3$ has also been used to synthesize highly efficient composites along with g-$C_3N_4$ **[118].** The formation of CoP/g-$C_3N_4$ and CoO/g-$C_3N_4$ hybrids was also studied and it was found that the $H_2$ evolution efficiency increased drastically in both the cases and type II heterojunctions were formed here as well **[119-120].** *Ran et al.*

synthesized 2D/2D van der Waals heterojunction using Phosphorene/g-$C_3N_4$, this was found to be a type-I heterojunction **[121].** g-$C_3N_4$/graphene composites have also been found efficient in the photodegradation of dyes due to the formation of a Schottky junction between g-$C_3N_4$ and graphene **[122-123].** The $SiO_2$/g-$C_3N_4$ heterojunction has shown about 5 (4.18) times higher efficiency that bare g-$C_3N_4$ in the photodegradation of RhB **[124].** $SnS_2$ has also been an efficient compound for heterojunction formation with g-$C_3N_4$. *Di et al.* formed a $SnS_2$/g-$C_3N_4$ direct Z-scheme heterojunction for $CO_2$ reduction **[125].** Other such composites formed include $MnO_2$/monolayer g-$C_3N_4$ (Z-scheme) **[126]**, $Ta_2O_5$/g-$C_3N_4$ **[127]]**, α-$Fe_2O_3$/g-$C_3N_4$(Z-Scheme) **[128]**, and $C_{60}$/g-$C_3N_4$ (Type I) **[129].**

**g-$C_3N_4$ as Sensors**

g-$C_3N_4$ based sensors can be classified broadly into two major types depending upon the sensing mechanism: (a) Fluorescent sensor and (b) Electrochemical sensor.

**<u>(a) Fluorescence Sensors:</u>**

In case of fluorescence sensors, fluorescent parameters such as intensity, wavelength, anisotropy, or lifetime of the probe change in the presence of the ions to be detected**.** Generally, fluorescence-based sensors adopt three different strategies: (a) fluorescence quenching (turn-off), (b) fluorescence enhancement (turn-on), and (c) fluorescence resonance energy transfer (FRET) **[130].** However, previous sensing probes had major drawbacks such as easy oxidation, toxicity, photobleaching, high cost etc. g-$C_3N_4$ has proved to be a good candidate as a fluorescent sensor since it is non-toxic, non-metallic, readily available and very robust against photobleaching. Synthesis of g-$C_3N_4$ in smaller dimensions such as QDs, nanosheets, nanorods further enhances the sensing efficiency due to increased surface area. g-$C_3N_4$ based sensors have found applications as biosensors as well. *Das et al.* observed the temperature dependent photoluminescence in g-$C_3N_4$ implying its potential application as low temperature sensors **[17].** *Shrivanand et al.* prepared a highly selective ON-OFF fluorescent sensor based on carboxyl rich g-$C_3N_4$ for the detection of $Hg^{2+}$ and $CN^-$ **[131].** The sensors for the detection of a single element or compound can be brought to detect more complex compounds. For example, Rong *et al.* used g-$C_3N_4$ nanosheets as a fluorescent switch for the detection of Chromium (VI) and further used it for sensing ascorbic acid. The limit of detection for chromium and ascorbic acid were 0.15 μM and 0.13 μM respectively **[132].** *Liu et al.* used ratio-metric fluorescence sensing strategy for the detection of $H_2O_2$ and based on the conversion of glucose into $H_2O_2$ through the catalysis of glucose oxidase they further used the sensor for the detection of glucose **[133].** *Duan et al.* synthesized a novel "on-off-on" fluorescent sensor for 6-Thioguanine (LOD= 65 nM) and $Hg^{2+}$ (LOD = 37 nM) based on g-$C_3N_4$ nanosheets **[134].** *Zhang et al.* synthesized g-$C_3N_4$ nanosheet-$MnO_2$ sandwich nanocomposite Turn-ON fluorescent sensor for the detection of glutathione (GSH) in aqueous solutions as well as living cells. Under optimum conditions the detection limit of 0.2 μM for GSH in aqueous solutions was reached **[135].** *Hu et al.* synthesized $Fe_3O_4$/g-$C_3N_4$/HKUST-1(Copper benzene-1,3,5-tricarboxylate) composites for the detection of ochratoxin-A with detection limit of 2.57 ng/mL **[136].** *Han et al.* used FRET based fluorescent sensor for the detection of riboflavin. Using this sensor, riboflavin can be detected from complex systems such as milk **[137].**

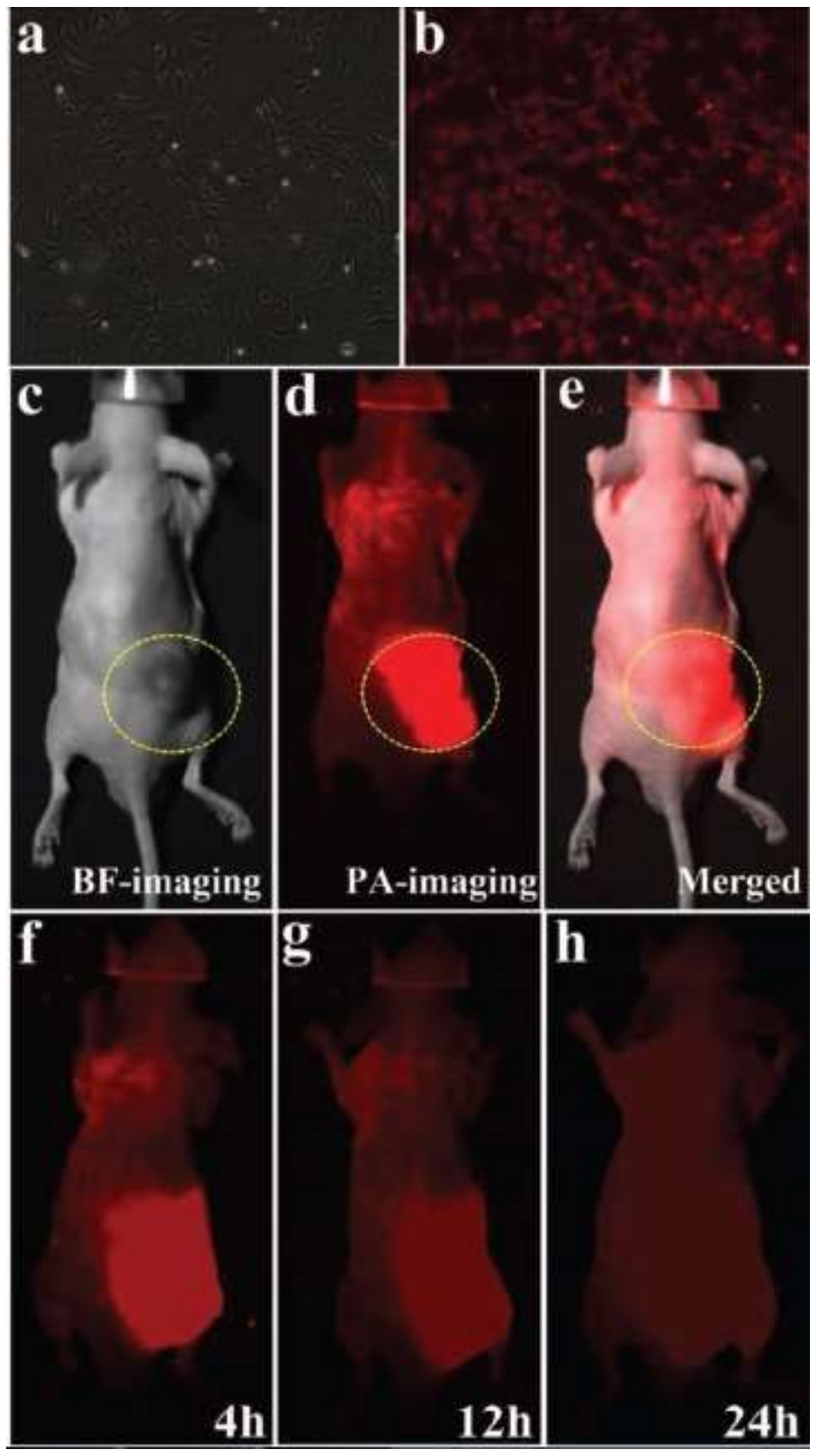

**Fig. 25** In vitro and in vivo bio-imaging of P-g-$C_3N_4$ QDs. (a) Bright-field microphotograph and (b) fluorescent microphotograph of OCM-1 cells incubated with P-g-$C_3N_4$ QDs-3. (c) Bright-field microphotograph of the nude mice tumor. (d) Photoacoustic image of tumor-bearing mice incubated with P-g-$C_3N_4$ QDs-3. (e) Bright-field and photoacoustic image of tumor-bearing mice incubated with P-g-$C_3N_4$ QDs-3. Yellow circles highlight the tumor site. Photoacoustic images of mice (f) 4, (g) 12, and (h) 24 h after the intratumoral injection of P-g-$C_3N_4$ QDs-3 (100 μL, 0.1 mg $mL^{-1}$) **[18].**

The fluorescent sensing capability of g-$C_3N_4$ and its composites also makes it a suitable material for bio-imaging applications. *Zhang et al.* implied the use of water soluble g-

$C_3N_4$ nanosheets for bioimaging as early as in 2013, due to the biocompatibility, non-toxicity, high stability and good quantum yield **[31].** Later on in 2014, another group *Zhang et al.* used $MnO_2$/g-$C_3N_4$ nanosheets for the intracellular imaging of glutathione **[135].** *Wu et al.* used phosphorous doped g-$C_3N_4$ QDs and recorded their bioimaging capabilities by incubating them with the OCM-1 cells as shown in Fig.25 **[135].**

**Table-2** Use of g-$C_3N_4$ as sensing material for various analyte.

| Target Analyte | Sensor Type | Material | Detection Limit | Sample Application | Reference |
|---|---|---|---|---|---|
| Various | Ratiometric (ECL, FL, PEC) | g-C3N4 composites | As low as 59 aM (RdRp gene) | Biosensing, food, environment | Xu, 2025 **[138]** |
| Organophosphorus pesticides | Fluorescence + Colorimetric | Cu2+ decorated g-C3N4 NS | 6.8 nM (FL); 1.2 nM (Colorimetric) | Cabbage, Tap Water | Chen, 2021 **[139]** |
| Chlortetracycline | Fluorescence (turn-on, AIE) | Citrate-modified g-C3N4 nanodots | 13 nM | Chicken, feed, cells | Y. X. Wu, 2024 **[140]** |
| Carboxin | Chemiluminescence | g-C3N4 Quantum Dots | Not specified | Water | Y. Zhou, 2024 **[141]** |
| Cu2+ | Fluorescence (turn-off) | Chitosan-modified g-C3N4 | 9 nM | Aqueous | Z. Li, 2024 **[142]** |
| Pb2+, Hg2+, Cd2+, etc. | Electrochemical | g-C3N4 nanocomposites | Depends on method | Aqueous environment | Yin, 2025 **[143]** |
| O2, Temp | Photoluminescence | g-C3N4 (urea and melamine based) | Not specified | Environmental monitoring | Kokkotos, 2025 **[144]** |
| $Pb^{2+}$ | Fluorescent aptasensor (FRET) | $g-C_3N_4$ nanosheets + AuNPs + Aptamer | 0.8 μg/mL | Environmental samples | L. Jia, 2025 **[145]** |
| Glutathione (GSH) | Ratiometric Fluorescence | $g-C_3N_4$ NSs + Ag NCs | 0.157 μM | Serum, tablets, cells | Sun, 2025 **[146]** |
| $H_2O_2$ | Fluorescence quenching + Colorimetric | W-doped $g-C_3N_4$ | 8 nM (FL), 20 nM (Colorimetric) | Biological/General | Ahmed, 2021 **[147]** |
| $H_2O_2$, Glucose | Ratiometric Fluorescence | $g-C_3N_4$ NS + CuNCs + $Ce^{3+}$ | 0.6 μM ($H_2O_2$), 0.48 μM (Glucose) | Milk, Serum | Mei, 2022 **[148]** |
| Chloride ions | Potentiometric | $g-C_3N_4$/AgCl nanocomposite | 0.4 μM | Water samples | Alizadeh, 2022 **[149]** |
| Histamine | Fluorescence (quenching) | $g-C_3N_4$ (KCl/NaCl assisted) | 0.61 mg/kg | Fish | Lin, 2022 **[150]** |
| Procalcitonin (PCT) | Electrochemical (non-enzymatic) | $g-C_3N_4$ NS + peptide | 0.11 fg/mL | Serum | Liu, 2022 **[151]** |

**Electrochemical Sensor**

The next type of sensor is the electrochemical sensor which also includes the photoelectrochemical sensors as well. The electrochemical sensors are further of two types: static and dynamic. Static sensors are used when there is no stream flow between the electrodes and the analyte and the sensing is governed by diffusion and Brownian motion. The dynamic sensors on the other hand are characterized by the current flow between the electrodes and the analyte as a result of redox reactions going on **[152].** Photoelectrochemical sensing refers to the influence of the interaction between the recognition element and the analyte on the photocurrent signal, which involves the charge and energy transfer of the PEC reaction between the electron donor /acceptor and the photoactive material upon light irradiation. N-doped carbons have received considerable attention due to the strong electron donor nature of nitrogen which enhances the collective π-bonding, leading to the improved stability, electron transfer rate and hence durability of the carbon supports during electrocatalytic processes **[153].** *Zhou et al.* synthesized chemically modulated g-$C_3N_4$ by exfoliating bulk carbon nitride (CN) of different polymerization degrees and developed highly selective ECL sensors for rapid detection of multiple metal-ions such as $Cu^{2+}$, $Ni^{2+}$ and $Cd^{2+}$. Electrochemiluminescence is a special form of chemiluminescence in which the light-emitting chemiluminescent reaction is preceded by an electrochemical reaction **[154].** *Raj kumar et al.* used sulfur doped g-$C_3N_4$ as electrode for electrochemical sensing of 4-Nitrophenol. They obtained a wide linear response range from 0.05–90 μM and a relatively low detection limit (0.0016 μM) **[155].** *Zou et al.* used sulfur doped g-$C_3N_4$ on glassy carbon electrode for lead ion ($Pb^{2+}$) sensing application **[156].** *Zhang et al.* synthesized ultrathin g-$C_3N_4$ modified glassy carbon electrode for $Hg^{2+}$ detection having a detection limit of 0.023 μg/L (S/N = 3) **[157].** *Cao et al.* synthesized $SnO_2$/g-$C_3N_4$ nanocomposites via calcination method for high performance ethanol gas sensors. The high surface area (132.5 $m^2g^{-1}$) of the composite made it even more efficient **[158].** *Zhai et al.* used ZnO/g-$C_3N_4$ composite for UV light assisted ethanol gas sensing **[159]** whereas *Zhang et al.* synthesized α-$Fe_2O_3$/g-$C_3N_4$ composite for ethanol sensing **[160].** *Hang et al.* synthesized graphene/g-$C_3N_4$ for $NO_2$ sensing **[161].**

In case of biosensors as well, there has been a significant amount of g-$C_3N_4$ composites for the detection of multiple elements. *Dong et al.* synthesized a dual-electrodes array modified by CdS@g-$C_3N_4$ p-n junction heterojunction for detecting prostate specific antigen (PSA). Here, the sensor was a photoelectrochemical sensor **[162].** *Tian et al.* used just g-$C_3N_4$ nanosheets as a highly efficient electrocatalyst for the reduction of $H_2O_2$ and further for glucose sensing. The sensor had a detection limit of 45 μM (SNR = 3) **[163].** *Lv et al.* synthesized 0-dimensional/ 2-dimensional (0D/2D) CQD/g-$C_3N_4$ nano-heterostructures for the sensitive photoelectrochemical (PEC) immunoassay of cancer-related biomarkers (prostate-specific antigen (PSA)) coupling with the copper nanoclusters (CuNCs). The PCE had a dynamic linear range of 0.02 – 100 ng $mL^{-1}$ and a limit of detection (LOD) of 5.0 pg $mL^{-1}$ **[164].** *Zou et al.* used 2D g-$C_3N_4$/CuO nanocomposites for dopamine detection. The response of the sensor was linear over the range of $2.00\times10^{-9}$ to $7.11\times10^{-5}$ mol $L^{-1}$ with a detection limit of $1.00\times10^{-10}$ mol $L^{-1}$ **[165].** *Liu et al.* developed a photoelectrochemical sensor for the detection of adenosine using CdS/PPy/g-$C_3N_4$ nanocomposites (PPy- polypyrrole for more efficient charge separation). Under optimal conditions, the PEC aptasensor had a sensitive response to adenosine in a linear range of 0.3 nmol$L^{-1}$ to 200 nmol$L^{-1}$ with a detection limit of 0.1 nmol$L^{-1}$ **[166].** *Liu et al.* synthesized an enzyme (glucose oxidase) biosensor under visible light irradiation using $TiO_2$ nanosheet/g-$C_3N_4$ nanocomposite. This photoelectrochemical sensor (PCE) had a linear range of 0.005-16 mM and a 0.01 mM glucose detection limit **[167].** *Cao et al.* detected glycophosate using photoelectrochemical sensor based on Cu-BTC MOF/g-$C_3N_4$ nanosheet (BTC-benzene-1,3,5-tricarboxylic acid) schematically as shown in Fig. 26. It had a low detection limit of $1.3\times10^{-13}$ mol $L^{-1}$ and a wide detection range ($1.0\times10^{-12}$ - $1.0\times10^{-8}$ mol $L^{-1}$ and $1.0\times10^{-8}$ ~ $1.0\times10^{-3}$ mol $L^{-1}$) **[168].**

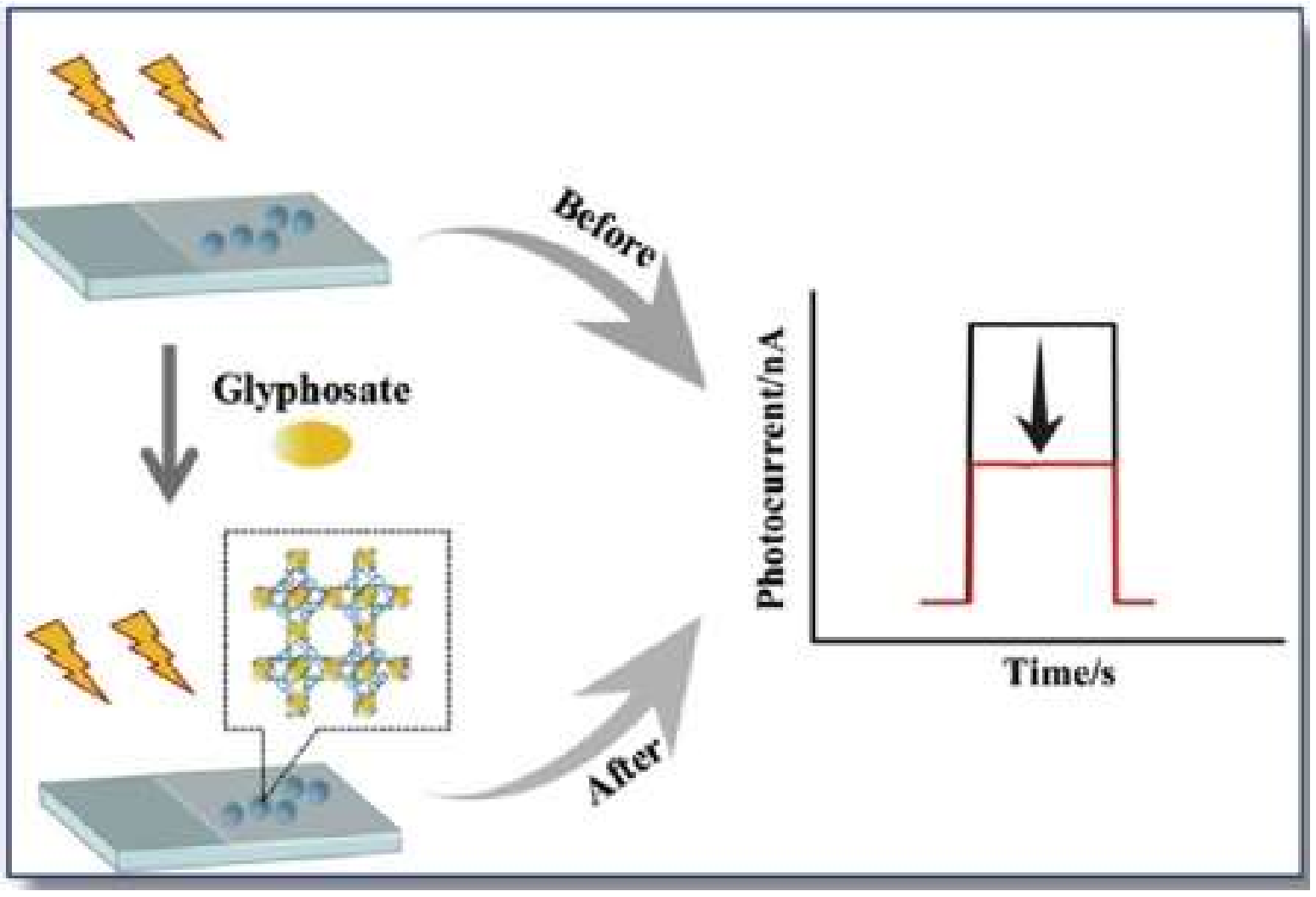

Fig. 26 The analytical principle of Cu-BTC/CN-NS based electrochemical sensor [168].

**Li-ion Storage**

Batteries are something that has found applications in our day to day life with Li ion batteries being one of the most common rechargeable batteries. Graphene or its derivatives have been extensively used as anode material for Li-ion batteries due to their high conductivity and good chemical stability **[169].** Here, we will be getting into the application of g-$C_3N_4$ composites as electrode material for Li ion storage. In 2014, *Hou et al.* reported the use of N-doped graphene/porous g-$C_3N_4$ nanosheets supporting layered-$MoS_2$ hybrid as a potential Li-ion battery anode. The hybrid nanosheets exhibit superior electrochemical performance, including excellent cycling stability (maintaining 91% capacity after 100 cycles), a high rate capability (retaining 83% capacity from 50 $mAg^{-1}$ to 500 $mAg^{-1}$), and a considerably large capacity (achieving more than 800 $mAhg^{-1}$ at 100 $mAg^{-1}$) **[170].** *Li et al.,* in the next year synthesized $Zn_2GeO_4$/g-$C_3N_4$ hybrids via facile solution approach. They observed that the 2D g-$C_3N_4$ layers effectively accommodated the volume changes and prevented the formation of unstable solid electrolyte interface (SEI), while monodisperse $Zn_2GeO_4$ NPs prevented the ultrathin g-$C_3N_4$ layers from restacking, promoted lithiation/ delithiation kinetics and contributed to the high specific capacities. Fig. 27 shows illustration of pure g-$C_3N_4$ and $Zn_2GeO_4$/g-$C_3N_4$ hybrids for Li-insertion viewed from the a,c edge and b,d basal plan directions. The composites exhibited a high charge capacity of 1370 mA h $g^{-1}$ at 200 mA $g^{-1}$ after 140 cycles and an excellent rate performance of 950 mA h $g^{-1}$ at 1000 $mAg^{-1}$. **[171].** *Vo et al.* used hydrothermal method to produce $SnO_2$/g-$C_3N_4$ nanocomposites and observed that the nanocomposite showed an improved cycling performance as compared to the $SnO_2$ alone **[172].** *Huu et al.* recently used $MoS_2$/g-$C_3N_4$ composite and studied their properties as anode material for Li storage. They observed that the composite had high capacity and stable cycling performance at 1 C ($Ag^{-1}$) with a reversible capacity of 1204 $mAhg^{-1}$ for 200 cycles. This was due to the role of g-$C_3N_4$ as a supporting material for the accommodation of volume change and improved charge transport **[19].** *Liu et al.* in the same year, synthesized $Li_4Ti_5O_{12}$/g-$C_3N_4$ composite. The composite showed excellent cycling stability with a maintained discharge capacity of 150.8 mAh/g at 0.5 C even after 502 cycles, corresponding to only 0.026 % capacity loss per cycle **[173].**

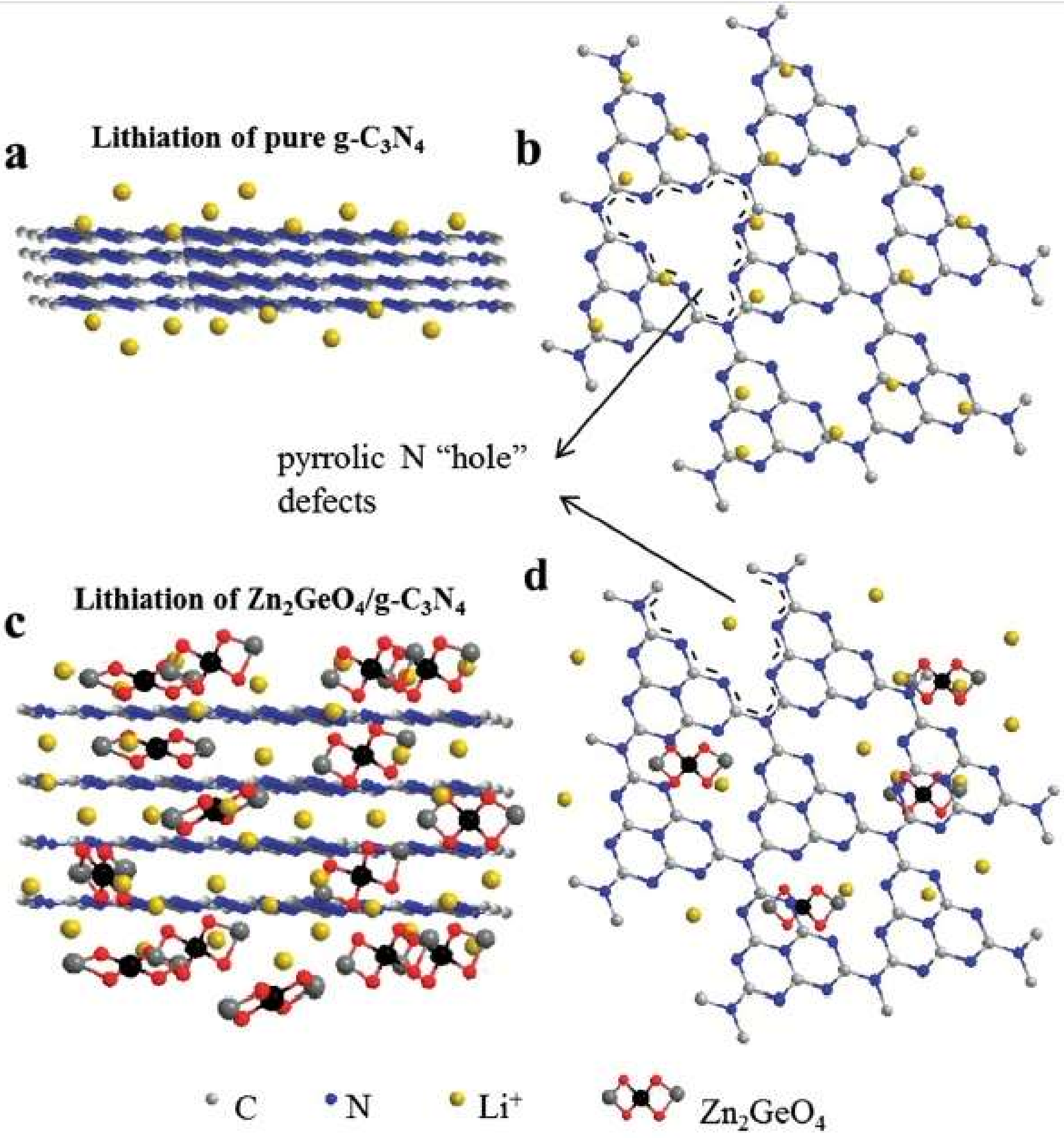

Fig. 27 Illustrations of pure a,b) g-$C_3N_4$ and c,d) $Zn_2GeO_4$/g-$C_3N_4$ hybrids for Li-insertion viewed from the a,c) edge and b,d) basal plan directions [171].

**Table-03: Use of graphitic carbon-nitride for Li-ion batteries**

| Sl. No. | material | Electrolyte | Initial Capacity (mAhg$^{-1}$) | Rate | Reversible capacity (mAhg$^{-1}$) | No of cycles | Reference |
|---|---|---|---|---|---|---|---|
| 1 | $C_3N_4$/NRGO/$MoS_2$ hybrid nanosheets | 1.0M $LiPF_6$ in ethylene carbonate/ethyl methyl carbonate (40:60 v/v) | 938 mAh g-1 | 100 mA g-1 | 855 | 100 | *Hou et al.(2014)* **[170]** |
| 2 | $Zn_2 GeO_4$ /g-$C_3 N_4$ | 1 M $LiPF_6$ in a mixture of ethylene carbonate (EC) and dimethyl carbonate (DMC) (v/v = 1:1). | 1370 mA h g −1 | 200 mA g −1 | 1304 | 100 | Li et al. (2015) **[171]** |
| 3 | $SnO_2$ nanosheets/g-$C_3N_4$ | 1 M $LiPF_6$ in ethylene carbonate (EC): ethyl diethyl carbonate (DEC): dimethyl carbonate(DMC) (3:3:4 volume ratio) | 733 | - | - | 60 | *Vo et al. (2017)* **[172]** |
| 4 | $MoS_2$/g-$C_3N_4$ | 1 M $LiPF_6$ in ethylene carbonate:ethyl diethyl carbonate:dimethyl carbonate (3:3:4 volume ratio) | 2467 | 1C | 1204 | 200 | Huu et al. (2019) **[19]** |
| 5 | $Li_4Ti_5O_{12}$/g-$C_3N_4$ | 1 M $LiPF_6$/ethyl methyl carbonate (EMC) + ethylene carbonate (EC) + dimethyl carbonate (DMC) (1:1:1 in volume) solution | 173.7 | - | 150.8 | 502 | *Liu et al. (2018)* **[173]** |
| 6 | Layered g-C3N4@Reduced Graphene Oxide | ethylene carbonate and dimethyl carbonate mixture (1:1) | 1476.1 mA h $g^{-1}$ | 595.1 mA h $g^{-1}$ | 949.4 mA h g−1 | 50 | Wang et al. **[174]** |
| 7 | $SnS_2$/g-$C_3N_4$/graphite | fluoro ethylenecarbona | 536.5 mA h | - | 470 mA h $g^{-1}$ | 30 | Zuo et al. **[175]** |
| | | te (FEC) | $g^{-1}$ | | | | |
| 8 | $Li_{1.2}$ $Mn_{0.54}$ $Ni_{0.13}$ $Co_{0.13}$ $O_2$ @g-$C_3N_4$ | - | 207.7 | 1C | 163.6 | 300 | F.-F. Wang et al. (2023) **[176]** |
| 9 | Graphitic carbon nitride | $Li_6PS_5Cl$ (LPSC). | 760 | 0.2C | 579 | 150 | Y. Zhao et al. (2024) **[177]** |
| 10 | Porous activated carbon integrated carbon nitride nanosheets a | 1 M LiTFSI | 1389 | 0.1C | 655 | 500 | J. Joseph et al. (2024) **[178]** |
| 11 | N and P dual-doped hollow carbon fibers/graphitic carbon nitride | 1 M $LiPF_6$ in ethylene carbonate (EC)/dimethyl carbonate (DMC)/ethyl methyl carbonate with a volume ratio of 1:1:1. | 1199 | $1Ag^{-1}$ | 1030 | 1000 | H. Tao et al. (2017) **[179]** |
| 12 | CuCo carbonate hydroxide nanowires/graphitic carbon nitride composites | 1 M $LiPF_6$ in a mixture of ethylene carbonate (EC)/dimethyl carbonate (DMC) (EC/DMC; 1:1 M ratio) | 1373 | 100 mA $g^{-1}$ | 394 | 100 | M.R. Raj et al. (2024) **[180]** |

**Supercapacitors**

g-$C_3N_4$ has been a strong potential candidate for supercapacitors. This is due to its high nitrogen content which provides it with more active reaction sites, increased electron donor/acceptor characteristics, enhanced charge transfer efficiency and large pseudo capacitance. *Shi et al.* reported $Ni(OH)_2$ hybridized g-$C_3N_4$ had a specific capacitance of 505.6 F/g at a current density of 0.5A/g which was higher than that of pure $Ni(OH)_2$ **[181].** *Dong et al.* also used $Ni(OH)_2$/g-$C_3N_4$ composite which exhibited higher specific capacities (1768.7 F/g at 7 A/g, 2667 F/g at 3 mV $s^{-1}$, respectively) compared to $Ni(OH)_2$ aggregations (968.9 F/g at 7 A/g) and g-$C_3N_4$ (416.5 F/g at 7 A/g), as well as better cycling performance (~ 84% retentions after 4000 cycles) **[182].** *Guo et al.* prepared high-performance electrode material for supercapacitors using porous g-$C_3N_4$ nanosheets/$NiCo_2S_4$ nanoparticles. The electrochemical characterizations showed that the as-prepared hybrid composite delivers excellent electrochemical properties, exhibiting a high capacitance (1557 F/g at current density of 1 A/g) and excellent cycling stability (only 7.4 % loss after 10000 cycles). This was attributed to the uniform distribution of $NiCo_2S_4$ on the surface of porous g-$C_3N_4$ and functioning as conductive linkers between g-$C_3N_4$ nanosheets, which ultimately improve the electrical conductivity of the hybrid composite for supercapacitor electrode **[183].** *Chen et al.,* synthesized PEDOT/g-$C_3N_4$ (PEDOT: poly (3,4-ethylenedioxythiophene): poly (styrenesulfonate)) as binary electrode material for supercapacitors. The specific capacitance of the composite was found to be 137 F/g in $H_2SO_4$ and 200 F/g in $Na_2SO_4$ at a current density of 2 A/g, respectively. Also more than 89 % and 96.5 % of capacitance were retained in $H_2SO_4$ and $Na_2SO_4$ respectively **[184].** *Chang et al.* used sandwich-like $MnO_2$/g-$C_3N_4$ nanocomposite electrode for supercapacitor. The composite enhanced the supercapacitor performance exhibiting high specific capacitance of 211 F/g at a current density of 1 A/g, with good rate capacity and cycling stability **[20].** *Dong et al.* used g-$C_3N_4$/PPy (polypyrrole) nano-composites as electrode material. The obtained nanocomposite exhibited large specific surface area and high specific capacitance, cyclic reliability and rate capability characteristics. The specific capacitance reached 471 F/g at current density of 1 A/g, remaining > 80 % of its original value with cycling up to 1000 times or at the current density of 20 A/g **[185].** *Sun et al.* further synthesized $MnO_2$/g-$C_3N_4$ and PPy (polypyrrole) composite and noted that the composite exhibited a high specific capacitance of 274 F/g at a current density of 2 A/g, which was 1.88 times than that of the $MnO_2$@PPy nanocomposite when used as electrodes. The $MnO_2$/g-$C_3N_4$@PPy nanocomposite electrode also possessed outstanding cyclic stability with 95 % capacitance retention after 1000 cycles **[186].** *Raghupathi et al.* used g-$C_3N_4$ doped MnS as electrode material and observed that the existence of spherical morphology and the lone pair of electron present in the graphitic carbon nitride improved the conductivity as well as supercapacitive behavior of the material leading to high specific capacitance of 463.32 F/g at 10 mV/s and 248 F/g at 1 A/g. The capacity retention of 98.6 % was observed after 2000 cycles **[187].** *Sanati et al.* reported the high capacitance and cyclic stability of CoAl-layered double hydroxide (CoAl-LDH) nanoflowers decorated with g-$C_3N_4$ nanosheets as an electrode material for supercapacitor applications. They demonstrated that adding various amounts of g-$C_3N_4$ nanosheets to the LDHs promotes its electrochemical properties by combining the redox reactivity of the LDH host with the considerable electronic conductivity of the g-$C_3N_4$. The nanocomposite presented a high specific capacitance of (343.3 F/g) at 5 A/g and presented 93% of its initial capacitance after 6000 cycles **[188].** Warsi *et al.* investigated g-$C_3N_4$/$Cd_2SnO_4$ as an electrode material for supercapacitor applications **[189]**. In a three-electrode configuration using 0.1 M $H_2SO_4$ as the electrolyte, the g-$C_3N_4$/$Cd_2SnO_4$ (synthesized at 500 °C for 2 hours) electrode exhibited a high specific capacitance of 601 $Fg^{-1}$ at a current density of 0.5 $Ag^{-1}$. The material also demonstrated excellent cycling stability, retaining 94.69 % of its initial capacitance after 5000 charge–discharge cycles at 5.0 $Ag^{-1}$. In a two-electrode setup, electrochemical analysis revealed a maximum specific capacitance of 351.45 $Fg^{-1}$, along with a power density of 250 $Wkg^{-1}$ at an energy density of 48.81 $Whkg^{-1}$, indicating strong potential for energy storage applications. Recently, in 2025 P. Chaluvachar et al. have presented a review on emerging role of graphitic carbon nitride in advanced supercapacitors **[190].**

**Table-4: Comparative analysis of graphitic carbon-nitride based supercapacitors**

| S. No. | Material | Synthesis route | Current density ($Ag^{-1}$) | Specific capacitance ($Fg^{-1}$) | Electrolyte | Energy density ($Whkg^{-1}$) | Power density ($Wkg^{-1}$) | Capacitance Retention | Reference |
|---|---|---|---|---|---|---|---|---|---|
| 1 | g-C3N4/Ni(OH)2 | Hydro-thermal | 0.5 | 505.6 | 6M KOH | 17.56 | 125.43 | 71.5% after 1000 cycles | Shi et al (2015) **[181]** |
| 2 | g-C3N4/Ni(OH)2 | - | 7 | 1768.7 | 6M KOH | 43.1 | 9126 | 72% after 8000 cycles | *Dong et al. (2017)* **[182]** |
| 3 | g-C3N4 nanosheets | multistep thermal treatment process | 1 | 1557 | 6M KOH | - | - | 92.6% after 10000 cycles | *Guo et al. (2017)* **[183]** |
| 4 | PEDOT/g-C3N4 | layer-by-layer assembly | 2 | 137 | H2SO4 | 17.5 | 5000 | 89% after 1000 | *Chen et al. (2015)* |
| | | method | | | | | | cycles | **[184]** |
| 5 | MnO2/g-C3N4 nanocomposite | soft chemical route at low temperature. | 1 | 211 | 0.5 M Na2SO4 | - | - | 100% after 1000 cycles | *Chang et al. (2017)* **[20]** |
| 6 | $C_3N_4$/PPy (polypyrrole) nano-composites | chemical oxidation method 6M KOH | 1 | 471 | 6M KOH | 23.7 | 14000 | >80% after 1000 cycles | *Dong et al. (2019)* **[185]** |
| 7 | MnO2/g-C3N4@polypyrrole (PPy) ternary nanocomposite | facile soft chemical routes | 2 | 274 | 0.5 M Na2SO4 | - | - | 95% after 1000 cycles | *Sun et al. (2018)* **[186]** |
| 8 | g-$C_3N_4$ doped MnS | sol-gel technique | 1 | 248 | 1 M Na2SO4 | - | - | 98.6% after 2000 cycles | *Raghupathi et al. (2019)* **[187]** |
| 9 | CoAl-LDH/g-C3N4 nanosheet composite | solvothermal method | 5 | 343.3 | 2M KOH | 61.15 | 13994.66 | 93% after 6000 cycles | *Sanati et al. (2019)* **[188]** |
| 10 | g-C3N4/Cd2SnO4 | mechanochemistry procedure | 0.5 | 601.20 | 0.1 M H2SO4 | 48.81 | 250 | 94.69% after 5000 cycles | Warsi *et al. (2025)* **[189]** |

**Electronic Devices**

g-$C_3N_4$ having its band-gap in the visible range and its tenability with doping, defect engineering or capability to form heterojunction is one of the suitable materials for electronic devices. In addition to these, robust & non-metallic nature and large light absorbing capabilities makes g-$C_3N_4$ as preferred material for electronic devices. *Zhao et al.* studied the photogenerated charge separation of g-$C_3N_4$/$TiO_2$ hybrid and found that due to the built-in electric field effect, the charge separation in g-$C_3N_4$/$TiO_2$ increased drastically **[112].** Liu Hongji et al. described a kind of metal-free multipurpose platform for NIR (808 nm) imaging-guided combination photo-chemotherapy made of g-$C_3N_4$ QDs embedded in carbon nanosheets (CNQD-CN) **[191]**. *Zhou et al.* on the other hand showed that the graphitic carbon nitride quantum dots CNQDs in the exposure 150 W of Xe lamp, underwent photodoping and turned from an insulator into a photoconductor. The photocurrent of the CNQDs photoconductor was already profound compared to the dark current, rapidly photo responsive in several seconds and highly reproducible by periodical light on and off. They deposited robust CNQDs film on a gold interdigital electrode with adjacent intervals of about 150 μm **[192].** *Sun et al.* studied the photoelectrochemical behavior of g-$C_3N_4$/ZnS/CuS and observed that the photocurrent of the g-$C_3N_4$/ZnS/CuS heterojunction photoconductor was profound compared to the dark current, having a rapid photo response time of several micro-seconds. The significantly highest photocurrent in the composite suggesting that the ternary heterojunctions have lowest recombination rate and most efficient charge separation **[193].** *Ye et al.* fabricated $MoS_2$/S-doped g-$C_3N_4$ heterojunction film which offered an enhanced anodic photocurrent of as high as $\sim 1.2 \times 10^{-4}$ A/$cm^2$ at an applied potential of +0.5 V vs

Ag/AgCl under the visible light irradiation **[194]**. *Lai et al.* in the same year reported g-$C_3N_4$ nanosheet/Graphene as hybrid phototransistors. These hybrid devices showed a high responsivity of the order $10^3$ $AW^{-1}$ and a gain of $10^4$ under UV illumination. The schematic of this device is shown in Fig. 28(a) and the transfer characteristic curve is shown in Fig. 28(b). This enormous photo response of the hybrid device was related to an efficient charge transfer from g-$C_3N_4$ nanosheets to graphene **[195].** *Velusamy et al.* made a flexible organic/inorganic 2D g-$C_3N_4$/$MoS_2$ photodetectors with tunable composition and photodetection properties. The photodetectors showed broadband photodetection suitable for both the UV and visible spectrum with good responsivity, specific detectivity and reliable and rapid photo-switching characteristics **[196].** *Prakash et al.* developed a multifunctional hybrid photodetector based on g-$C_3N_4$/p-Si heterojunction. The built-in field at the n–p junction allowed the device to operate over ultrabroadband region from 250 to 1650 nm in self-powered mode. The device shows a novel binary photo switching in response to OFF/ON light illumination at small forward bias ($\leq$ 0.1 V) covering 250–1350 nm. At zero bias, the device displays an extremely high ON/OFF ratio of $\approx 1.2 \times 10^5$ under 680 nm (49 □W $cm^{-2}$) illumination. The device also showed an ultrasensitive behavior over the entire operating range at low light illuminations with highest responsivity (1.2 $AW^{-1}$), detectivity (2.8 $\times 10^{14}$ Jones), and external quantum efficiency (213 %) at 680 nm **[197].**

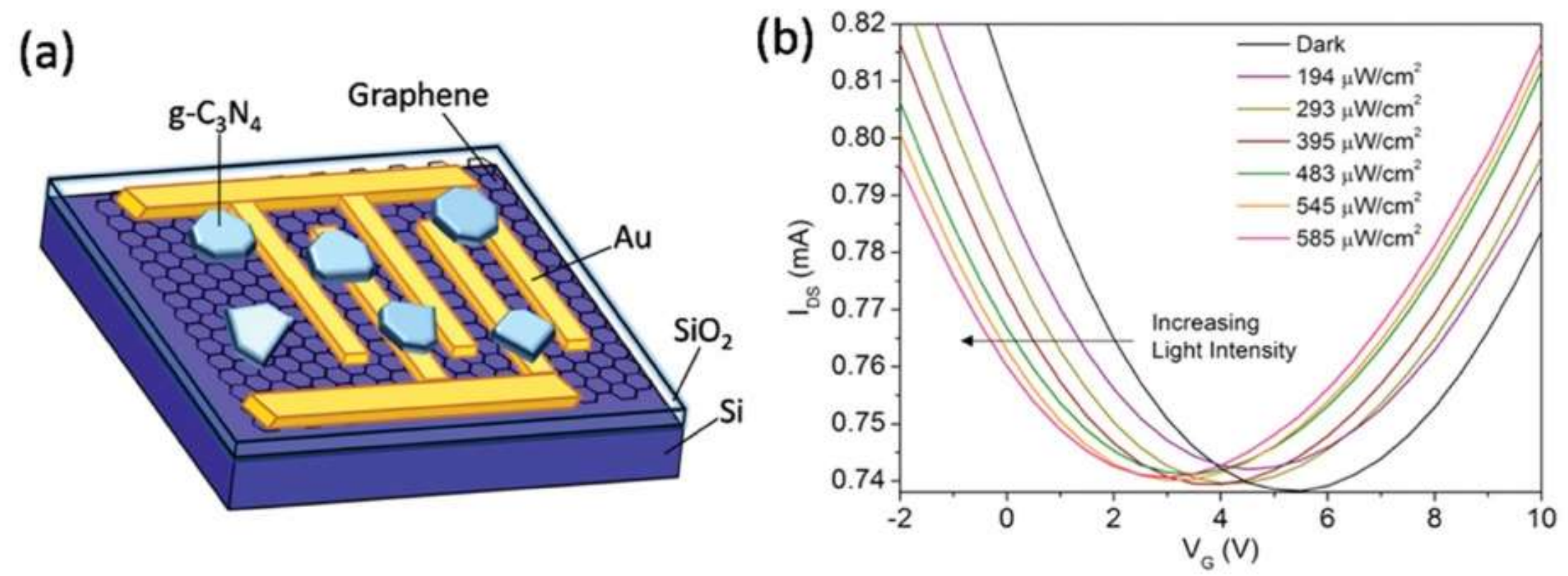

**Fig. 28** a) Schematic diagram of the device based on g-$C_3N_4$ and graphene. b) Transfer curves of the device at excitation wavelength is 370 nm under different light intensities having applied $V_{DS}$ of 0.5 V **[195].**

**Table-5** Summary of performance parameters of g-$C_3N_4$ based hybrid photodetectors.

| **Sl. No.** | **Active Materials** | **Growth Technique** | **Excitation wavelength** | **Responsivity (A/W)** | **EQE (%)** | **REFRENCES** |
|---|---|---|---|---|---|---|
| 1. | g-$C_3N_4$ /graphene | Thermal condensation | 370 nm | $4\times10^3$ | - | Lai Sin Ki, et al., 2016 **[195]** |
| 2. | g-$C_3N_4$ /Si | Drop-casting | 680 | 1.2 | 213 % | Prakash Nisha, et al., 2018 **[197]** |
| 3. | g-$C_3N_4$ /$MoS_2$ | Liquid Phase Exfoliation | 365 and 532 nm | 4 | - | Velusamy, Dhinesh Babu, et al., 2019 **[198]** |
| 4. | Au-g-$C_3N_4$ /CdS/ZnO | Pyrolysis | 530 nm | - | - | S. Pal et al., 2018 **[199]** |
| 5. | g-$C_3N_4$ /N-doped graphene | Polymerization | 365 nm and 488 nm | 0.59 mA/W and ~30 μA/W | - | X. Gan et al., 2018 **[200]** |
| 6. | CNT/ g-$C_3N_4$ nanosheets | spin coating method | 365 nm | 0.23 | 0.01-10 % | H. Fang et al., 2020 **[201]** |
| 7. | g-$C_3N_4$ /GaN | Hydrothermal treatment | 360-635 nm | 3 | $10^3$ | K. Sarkar et al., 2021 **[202]** |
| 8. | H-gCN (hydrogen substituted g-$C_3N_4$) 2D sheets | Spin-coated | 350-1100 nm | 0.34 | 59 | A. Ghosh et al., 2021 **[203]** |
| 9. | g-$C_3N_4$/Bi | Ultrasonication | 300–800 nm | $2.843\times10^{-3}$ | - | Y. Zhang et al., 2021 **[204]** |
| 10. | g-$C_3N_4$ /GaN nanowires | Spin-coated | 392 nm | 0.02 | - | M. Reddeppa et al., 2021 **[205]** |
| 11. | g-$C_3N_4$ film | Vapor Phase Transport | 355 | $5 \times 10^{-5}$ | $1.75 \times 10^{-2}$ % | Z. Liu et al. 2021 **[206]** |
| 12. | Ag/MWCNT / g-$C_3N_4$ nanosheets | Solution process | 534 nm | 0.9 A/W | 151 % | S. V. Manikandan et al., 2022 **[207]** |
| 13. | g-CN/Si | Vapor-phase transport-assisted condensation method | 515 nm | 133 | - | X. Chen et al., 2023 **[208]** |
| 14. | Sulfur-doped g-$C_3N_4$/GaN | Calcination process | 365 nm | 581 mA/W | 3.75 % | W. Song et al., 2024 **[209]** |
| 15. | g-$C_3N_4$/$SnSe_2$/H-$TiO_2$ | Impregnation and CVD | 370, 450, 520 nm | 2.742 A $W^{-1}$ | $9.21 \times 10^2$ % | X. Zhao et al., 2024 **[210]** |
| 16. | g-$C_3N_4$/CdS- | Spin coated | 2.16 μW/c | 0.39 | - | K. N. Das |
| | coated ITO/PET and Al foil | | m2 (Dark Condition) | V/mW | | et al., 2024 **[211]** |
| 17. | V4C3Tx MXene/protonated g-$C_3N_4$ | Spin Coating | 365 nm | - | - | P. M. A. Kenichi, et al. 2025 **[212]** |
| 18. | g-$C_3N_4$ | Calcination | 360 nm | 17.6 | 60.5 | S. Kumari et al., 2025 **[213]** |
| 19. | g-C3N4/p-Si | Drop-casting | 405 | 32 | $80 \times 10^2$ | R. K. Prasad et al. 2025 **[214]** |

**Table -6** Recently improved photovoltaic solar cells performance based on g-$C_3N_4$.

| **S. No.** | **Structure** | **PCE (%)** | **FF (%)** | **$J_{SC}$ (mA .cm$^{-2}$)** | **$V_{OC}$ (V)** | **Reference** |
|---|---|---|---|---|---|---|
| 1. | g-$C_3N_4$ added carbon | 14.34 | 60.1 | 23.80 | 1.00 | Yang, Zong Lin, et al., 2019 **[215]** |
| 2. | FTO/compact $TiO_2$/exfoliated g-$C_3N_4$ modified $MAPbI_3$/spiro-OMeTAD/Au | 15.80 | 62 | 23.2 | 1.1 | Wei Xiang feng, et al., 2019 **[216]** |
| 3. | Iodine doped g-$C_3N_4$ | 17.53 | 73 | 22.63 | 1.06 | Cao Wei, et al., 2019 **[217]** |
| 4. | FTO/ $TiO_2$/ g-$C_3N_4$ | 19.49 | 74 | 24.31 | 1.07 | Jiang Lu Lu, et |
| | | | | | | al., 2018 **[218]** |
| 5. | Perovskite / g-$C_3N_4$ | 18.09 | 75.96 | 22.47 | 1.06 | Li Zhen, et al., 2020 **[219]** |
| 6. | 2D g-$C_3N_4$ | 19.67 | 80.7 | 21.45 | 1.14 | Liu Zhou, et al., 2020 **[220]** |
| 7. | FTO/compact $TiO_2$ / g-$C_3N_4$ | 21.1 | 79 | 23 | 1.16 | Liao Jin Feng, et al., 2019 **[221]** |
| 8. | g-$C_3N_4$ QDs | 21.23 | 80.7 | 21.45 | 1.14 | Liu Pei, et al., 2020 **[222]** |
| 9. | g-$C_3N_4$ QDs doped $SnO_2$ | 22.13 | 78.3 | 24.03 | 1.18 | Chen Jinbo, et al., 2020 **[223]** |

**Future Outlook**

Due to versatile physicochemical properties and numerous applications, graphitic carbon nitride (g-$C_3N_4$) holds strong potential for application in future generation devices. It's one of the important materials of interest being metal free 2-dimensional semiconductor having high thermal stability, tunable bandgap (~ 2.7 eV) and biocompatibility. Additionally, being cost-effective material it has attracted researchers and industries working in the domain of sensors, energy storage and environmental remediation. In spite of reported number of techniques for synthesis of g-$C_3N_4$, method to synthesize g-$C_3N_4$ with controlled stoichiometric ratio and defined phase lacks in the literature. The stoichiometric ratio, phase and defect density controls its physical and chemical properties limiting its potential application. There has been ongoing efforts to dope it with various dopants like phosphorous, zinc, cobalt, nickel, gold, silver and magnetism to tailor its electronic, optical and photocatalytic properties.

g-$C_3N_4$ has been found to play transformative role in the domain of sensors. Its surface chemistry and electronic structure can be tailored for high sensitivity and specificity, which is crucial for detecting a wide range of optical, biological and chemical analytes. Recent research activities are focussed to enhance its conductivity and signal amplification through strategies such as heterojunction formation, defect engineering and doping. These modifications could significantly improve its performance in electrochemical and fluorescence-based sensors for medical diagnostics, environmental monitoring and food safety applications.

Environmental remediation is another crucial area where g-$C_3N_4$ shows immense potential. Its photocatalytic activity under visible light enables the degradation of pollutants, including pharmaceuticals, dyes and organic waste in water. There lies a great scope to further enhance its photocatalytic performance through composite formation with materials such as $TiO_2$, ZnO, reduced graphene oxide (rGO) and through introduction of dopants. Composite formation and introduction of dopants is expected to enhance the separation efficiency of photo generated charge carriers. These enhancements could enable g-$C_3N_4$ based decentralized, solar-powered water purification systems.

For energy storage applications, g-$C_3N_4$ is being explored as a catalyst for water splitting and $CO_2$ reduction. It processes required properties for clean hydrogen production and carbon mitigation. Structural modifications such as the integration of transition metals or co-catalysts, can further improve catalytic efficiency and stability. These advancements indicate towards future possibility of application of g-$C_3N_4$ to realize green hydrogen generation and carbon-neutral fuel technologies.

Membrane technologies also stand to benefit from the integration of g-$C_3N_4$. Its incorporation into polymeric matrices enhances the selectivity and mechanical strength of membranes used in ultrafiltration, oil-water separation and gas separation. Compared to traditional nanofillers, g-$C_3N_4$ offers better dispersion in organic matrices and superior environmental compatibility. The next phase of research are likely to focus on multifunctional membranes that combine separation, catalysis and disinfection properties, offering a holistic solution to complex industrial and municipal wastewater challenges.

In the realm of fuel cells and energy storage, g-$C_3N_4$ is gaining traction as a support material or even as a co-catalyst in direct methanol fuel cells (DMFCs). When combined with non-noble metals like palladium or iron, it enhances methanol oxidation reactions while resisting CO poisoning, a major limitation in conventional platinum-based systems. Furthermore, its high surface area and chemical tunability makes it suitable for developing high-performance supercapacitors and lithium-sulphur batteries.

Beyond these applications, g-$C_3N_4$ holds potential as a catalyst in organic synthesis and Fenton-like reactions. Its basic sites and ability to activate hydrogen peroxide makes it useful for green catalytic processes. These emerging applications highlight the adaptability of g-$C_3N_4$ across diverse chemical transformations, further underscoring its importance in future sustainable technologies.

In conclusion, the multifaceted capabilities of g-$C_3N_4$ position it as a cornerstone in the development of next-generation technologies for clean energy, environmental protection and advanced sensing. Overcoming current limitations, such as poor charge mobility and limited surface area, through innovative material engineering will be essential for unlocking its full potential. As research continues to evolve, g-$C_3N_4$ is set to become a key player in the global pursuit of sustainable and intelligent materials.

**Acknowledgment**

Dilip K. Singh thanks UGC-DAE CSR Indore (CRS/2021-22/01/358) and DST, Government of India (CRG/2021/002179; CRG/2021/003705; CRG/2023/002324) for financial support.

**Declaration of competing interest**

On behalf of all authors, the corresponding author states that there is no conflict of interest.